\documentclass[12pt, draftclsnofoot, onecolumn]{IEEEtran}
\usepackage{cite}
\usepackage{amsmath,amssymb,amsfonts}
\usepackage{algorithmic}
\usepackage[linesnumbered, ruled, vlined, noend]{algorithm2e}
\usepackage{graphicx,color}
\usepackage{float}
\usepackage{textcomp}
\usepackage{xcolor}
\usepackage{hyperref}
\usepackage{bm}
\usepackage{dsfont}
\usepackage{ctable}
\usepackage{comment}

\def\BibTeX{{\rm B\kern-.05em{\sc i\kern-.025em b}\kern-.08em
    T\kern-.1667em\lower.7ex\hbox{E}\kern-.125emX}}
\AtBeginDocument{\definecolor{tmlcncolor}{cmyk}{0.93,0.59,0.15,0.02}\definecolor{NavyBlue}{RGB}{0,86,125}}

\newtheorem{definition}{Definition}
\newtheorem{theorem}{Theorem}[section]

\newtheorem{proposition}[theorem]{Proposition}
\usepackage{subcaption}

\begin{document}

\title{Deep Reinforcement Learning for Uplink Scheduling in NOMA-URLLC Networks}

\author{\IEEEauthorblockN{Benoît-Marie~Robaglia\IEEEauthorrefmark{2},
Marceau~Coupechoux\IEEEauthorrefmark{2}, 
Dimitrios~Tsilimantos\IEEEauthorrefmark{1},
}

\IEEEauthorblockA{\IEEEauthorrefmark{2}LTCI, Telecom Paris, Institut Polytechnique de Paris}

\IEEEauthorblockA{\IEEEauthorrefmark{1}Advanced Wireless Technology Lab, Paris Research Center, Huawei Technologies Co. Ltd.}

\thanks{Corresponding author: Marceau~Coupechoux (email: marceau.coupechoux@telecom-paris.fr)}
\thanks{The work of B.-M. Robaglia and M. Coupechoux has been performed at the LINCS laboratory (lincs.fr).}}

\maketitle

\vspace{-1cm}
\begin{abstract}

This article addresses the problem of Ultra Reliable Low Latency Communications (URLLC) in wireless networks, a framework with particularly stringent constraints imposed by many Internet of Things (IoT) applications from diverse sectors. We propose a novel Deep Reinforcement Learning (DRL) scheduling algorithm, named NOMA-PPO, to solve the Non-Orthogonal Multiple Access (NOMA) uplink URLLC scheduling problem involving strict deadlines. The challenge of addressing uplink URLLC requirements in NOMA systems is related to the combinatorial complexity of the action space due to the possibility to schedule multiple devices, and to the partial observability constraint that we impose to our algorithm in order to meet the IoT communication constraints and be scalable. Our approach involves 1) formulating the NOMA-URLLC problem as a Partially Observable Markov Decision Process (POMDP) and the introduction of an {\it agent state}, serving as a sufficient statistic of past observations and actions, enabling a transformation of the POMDP into a Markov Decision Process (MDP); 2) adapting the Proximal Policy Optimization (PPO) algorithm to handle the combinatorial action space; 3) incorporating prior knowledge into the learning agent with the introduction of a Bayesian policy. Numerical results reveal that not only does our approach outperform traditional multiple access protocols and DRL benchmarks on 3GPP scenarios, but also proves to be robust under various channel and traffic configurations, efficiently exploiting inherent time correlations.\end{abstract}


\maketitle

\section{INTRODUCTION}
 
 \IEEEPARstart{N}{ew} high-demanding use cases pertaining to various industry sectors need to be addressed by the future generations of wireless networks\footnote{The work of B.-M. Robaglia and M. Coupechoux has been performed at the LINCS laboratory (lincs.fr).}. In particular, the Third Generation Partnership Project (3GPP) standard \cite{3gpp.38.913} has defined Ultra Reliable Low Latency Communications (URLLC) requirements for many Internet of Things (IoT) use cases such as smart grids, factory automation  and intelligent transportation to only name a few. A classical URLLC reliability requirement is for example to transmit a 32-byte packet with success probability $1-10^{-5}$ and with a latency deadline of $1$~ms~\cite{3gpp.38.913}. A deadline is said to be \emph{strict} if the packet is lost beyond this delay. URLLC requirements are particularly challenging on the uplink, i.e., from IoT devices to a central Base Station (BS), because the BS can acquire traffic and channel information only at the cost of a significant signalling load and delay; a problem related to \emph{partial observability} in control theory. In order to reduce latency and improve reliability, Non-Orthogonal Multiple Access (NOMA) is seen as a promising transmission technique, as it allows to schedule multiple users on the same time-frequency resource and to improve the spectral efficiency~\cite{saito2013non}. However, even  with NOMA, there is in practice a limited number of users sharing the same resource and the {\it user selection} issue adds to the complexity of the traditional many-to-one scheduling problem. In this context, we thus propose NOMA-PPO, a new Deep Reinforcement Learning (DRL) scheduling algorithm to solve the NOMA uplink URLLC scheduling problem with strict deadlines. 

\subsection{Related Work}

\subsubsection{Uplink URLLC access solutions}

Uplink access schemes for URLLC can be divided in two main groups, namely grant-based and grant-free protocols. Both approaches can be extended using NOMA or Deep Reinforcement Learning (DRL). 

In the first set, the scheduling of the devices is performed by the BS, see e.g.~\cite{Cuozzo22, Nomeir21}. Devices with a packet to transmit first send a scheduling request on the uplink. The BS then allocates uplink resources for the packet transmission. Uplink packets may include in their header some scheduling information (like the buffer status) to avoid the scheduling request step. In this case, a scheduling algorithm is required at the BS to meet the delay and reliability constraints without losing resources when a polled device has no packet to transmit. This is the baseline protocol adopted in 5G New Radio (NR)~\cite{3gpp.38.321}. The main drawback of the approach lies in the duration of the four-way handshake that may be incompatible with URLLC constraints. The advantage, though, is to avoid collisions between device transmissions. 

In the second set of access schemes, the handshaking is removed by allowing uplink transmissions to be grant-free (GF). This means that devices can transmit without an explicit command from the BS. We can further distinguish contention-free and contention-based GF access. In contention-free GF (also called semi-persistent scheduling), the BS pre-allocates periodic orthogonal uplink resources to the devices, so that there are no collisions~\cite{Feng19}. When a device has a packet to send, it waits for the next opportunity. This access scheme has been also adopted by 5G NR~\cite{3gpp.38.321}. Contention-free GF is however mostly adapted to periodic deterministic traffic, but becomes inefficient when the traffic is sporadic or probabilistic because resources may be lost, if there is no packet to be sent, or deadlines violated, when the packet arrival rate is suddenly higher. 

Several papers have studied contention-based GF, a family of protocols that are versions of Slotted Aloha (SA) enriched with smart retransmission schemes. Contrary to other approaches, uplink transmissions are indeed here subject to collisions. A typical example of this literature is the work presented in~\cite{elayoubi2019radio}, where authors adapt SA to URLLC and industrial IoT use cases by introducing retransmission schemes that depend on the traffic profile of the devices. In~\cite{mahmood2019uplink}, authors summarize the classical retransmission schemes: the K-repetition GF scheme, in which a pre-determined number of copies of the same packet are transmitted; the reactive GF scheme, in which devices receive a feedback from the BS for every transmission; and the proactive GF scheme, in which a packet is repeatedly sent until a positive acknowledgement is received. These protocols have been enhanced using NOMA~\cite{saito2013non} with the goal of better using the available resources and reduce the number of collisions for a given traffic load, see e.g.~\cite{mahmood2019uplink, tegos2020slotted, shahab2020grant} and references therein. However, with or without NOMA, all SA-based approaches suffer from high collision rates when the load or the number of devices increases~\cite{liu2020analyzing} and fail to take advantage of the various traffic patterns or channel conditions across the devices.

Recent advances in Deep Reinforcement Learning (DRL) \cite{mnih2015human} have been applied to solve several limitations in IoT systems \cite{chen2021deep} and are potential solutions for the aforementioned problems.
Several proposals use Deep Multi-Agent Reinforcement Learning (MARL) to model a user with a DRL algorithm in order to learn a transmission protocol in a decentralized manner in the context of dynamic spectrum access  \cite{chang2018distributive, xu2020application, tan2021cooperative}. Nonetheless, these solutions do not tackle the URLLC constraint with strict deadlines and do not take into account the potential of NOMA. The approach of \cite{yang2020deep} models the massive access problem by transforming the URLLC constraint into a data rate constraint and learns a transmission strategy in order to maximize the network energy efficiency using cooperative MARL. However, the authors do not consider strict deadlines and do not address the theoretical limitations of decentralized MARL like the non-stationarity during training.

At last, several strategies leveraging DRL have been put out to deal with the URLLC constraint in NOMA systems. The authors of \cite{ahsan2022reliable} propose Deep-SARSA to tackle the resource allocation problem at the BS for minimizing the error probability in uplink transmissions. Yet, the proposed solution does not take into account the packet arrival processes, assumes full observability of the system and does not impose strict deadlines. Additionally, the work of \cite{liu2021general} optimizes a NOMA based GF protocol with DRL. The authors use DRL to dynamically adjust the number of repetitions and radio resources in the proactive GF scheme. Nevertheless, the approach, which is based on SA, still suffers from a high collision rate as the load increases, is not designed for handling both deterministic and sporadic traffic and fails to take advantage of channel correlations. A very preliminary version of our work has been presented in~\cite{robaglia2021deep}. However, the proposed solution ignores NOMA, does not take into account channel correlations and is only adapted to probabilistic periodic traffic. 

In this paper, we consider a system in which the BS semi-blindly schedules the devices for their uplink transmissions, as it is done in grant-based access, however without the need for scheduling requests. Thanks to NOMA, the BS is able to poll multiple devices for a transmission in the same resource. We thus tackle a partially observable scheduling problem where the BS should strike a balance between acquiring scheduling information and avoiding excessive collisions. Our problem is characterized by two challenges, namely a combinatorial action space and a partially observable environment, that conventional DRL algorithms fail to handle.

\IEEEpubidadjcol

\subsubsection{DRL challenges for uplink URLLC}

First, allowing the BS to poll multiple devices in a frame drastically increases the action space. For $k$ devices, the decision maker needs to choose between $2^k$ actions, which is exponential in $k$. Few solutions have been proposed to address this problem in the literature. The most common one is proposed in \cite{dulac2015deep}. The authors' idea is to project the large discrete action space in a continuous action space and thus solve a continuous action RL problem with the traditional Deep Deterministic Policy Gradient algorithm \cite{lillicrap2015continuous}. However, this approach assumes that the discrete action space can be embedded in a continuous space, which is not straightforward for our Multiple Access (MA) problem. An alternative is the work of \cite{metz2017discrete}. The authors solve a high dimensional action space RL problem with a Recurrent Neural Network (RNN) to sequentially predict the action vector, one dimension after the other. Nevertheless, not only does this algorithm assume that we know how to order the action dimensions, but the Q-value estimated for the last dimension is very noisy, especially when $k$ is large. An extension of this paper is the Branching Dueling Q-Network (BDQ) \cite{tavakoli2018action}. The authors solve a RL problem with a $k$-dimensional action space using a dueling architecture where there is a value network common for all dimensions and $k$ advantage networks, one for every dimension. Yet, not only is this solution ill-suited to manage partial observability, but it also cannot account for any prior knowledge the agent might have regarding the dynamics of the environment.

Second, as the BS is not aware of the whole environment and takes decisions solely based on partial observations of the environmental state, our problem can be modeled by a POMDP~\cite{sondik1971optimal}. When observations are not Markovian, traditional RL algorithms work with history dependent policies, an approach that can be rapidly computationally intractable as the number of possible histories grows exponentially with the horizon. 
A way to alleviate this problem is to introduce \emph{belief states}, a probability distribution over the states, which is also a sufficient statistic for the past history and the initial state distribution. A POMDP can be then reformulated as a MDP in which the state space is the continuous belief state space. Traditional RL methods like Q-learning or policy gradient algorithms can finally be used on the resulting belief-MDP~\cite{kaelbling1998planning}.

Three main methods are proposed in the literature to derive or estimate a belief state: 1) the belief update formula \cite{kaelbling1998planning}, 2) a RNN \cite{hausknecht2015deep} and 3) a generative model \cite{igl2018deep}. However, all these methods suffer from major drawbacks. While the belief update formula requires the knowledge of the environment dynamics (transition and observation function), using a RNN or a generative model introduces a new layer of complexity since there are now two phases involved: the belief estimation and the computation of the optimal policy. Additionally, since Deep Neural Networks (DNNs) are black boxes, it is impossible to add any prior knowledge that the agent might have about the environment. Moreover, the learned beliefs are difficult to interpret and error might be propagated to the policy optimization phase.

An alternative to the belief state is the notion of \emph{information state} \cite{subramanian2022approximate} or \emph{internal state} \cite[Section 12.4.2]{wiering2012reinforcement}. The idea is to derive a function of the history which is a sufficient statistic for estimating the environmental state. However, learning such a sufficient representation of the history is difficult as it is often task-specific.

\subsection{Contributions and outline}

In this paper, we formulate the NOMA-URLLC problem as a partially-observable scheduling problem and solve it by proposing a DRL algorithm. Our contributions can be summarized as follows:
\begin{itemize}
    \item We formulate a general MA problem with the URLLC constraint, considering packets with strict deadlines and NOMA uplink communications as a POMDP. 
    \item We introduce the notion of \emph{agent state} in order to theoretically address the POMDP formulation. We show that the agent state is a sufficient statistic for the past observation-action history that allows us to 1) express past actions and observations in a compact way, and 2) convert the POMDP problem to an MDP and benefit from the convergence properties of the DRL algorithms. 
    \item We propose a DRL algorithm, \emph{NOMA-PPO}, that enhances the state-of-the-art algorithm PPO \cite{schulman2017proximal} with a branching policy network architecture in order to linearly manage combinatorial action spaces. This idea is inspired by the BDQ architecture \cite{tavakoli2018action} and extended to PG methods. In particular, this allows us to define Bayesian policies and incorporate prior information about the MA problem into the DRL agent \cite{titsias2018bayesian}. 
    \item We provide numerical evidence that our approach outperforms traditional MA and DRL benchmarks across 3GPP scenarios in terms of URLLC score, convergence speed, and fairness. Furthermore, we show that our algorithm is able to cope with different traffic models, a deterministic periodic and a probabilistic aperiodic traffic model in particular. Finally, our algorithm exhibits robustness against different channel configurations and demonstrates a successful exploitation of time-varying channel information.
\end{itemize}

In Section~\ref{section:system_model}, we define the system model. Section~\ref{section:problem_formulation} formulates the POMDP problem. Section~\ref{section:nomappo} presents the NOMA-PPO approach and finally, Section~\ref{section:experiments} exposes the simulations and numerical results. 

{\bf Notations:} For a finite set $X$, $\Delta(X)$ denotes the set of all probability distributions over $X$. The indicator function is denoted $\mathds{1}\{\cdot\}$, $diag(\cdot)$ is the diagonal operator that transforms a vector in a diagonal matrix and $\odot$ is the Hadamard product. The matrices are written in bold upper case and the vectors in bold lower case. $\langle \cdot \rangle$ refers to a tuple, $[\cdot]$ to the modulo operator and $\arg_B \min(S)$ returns a set of $B$ elements in $S$ having the lowest value (ties are broken at random). 

\section{System Model}\label{section:system_model}

\subsection{Network Model}

\begin{figure}[t]
    \centering
    \includegraphics[width=0.5\linewidth]{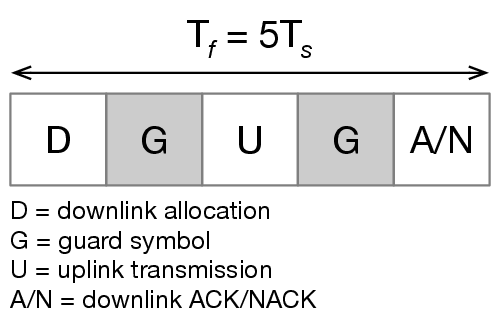}
    \caption{Radio frame structure.}
    \label{fig:slot}
\end{figure}

We consider a time-slotted wireless network of $K$ heterogeneous devices communicating with a BS over a wireless shared channel on the uplink. Every device has a single antenna and the BS is equipped with $n_a$ antennas. The time is divided into radio frames of duration $T_f$ and every frame is divided into five time-slots of duration $T_s$ (see Fig.~\ref{fig:slot}). This division represents the minimum time required for the processes of polling, transmitting and acknowledging. The time synchronization among all devices is performed by the BS using downlink signals. During the first slot of every radio frame, the BS is allowed to poll a number of devices for a potential uplink transmission, described by the vector $\boldsymbol{a}=(a_1, a_2, \dots, a_K) \in \{0,1\}^K$, where $a_k=1$ when the device $k$ is polled and $a_k=0$ otherwise. It also allocates orthogonal resources for uplink pilot transmissions from the polled devices. After a guard interval, a \emph{polled} device with at least a packet in its buffer becomes \emph{active} and transmits during the third slot. Its transmission includes a pilot signal for channel estimation, sent using the orthogonal resource allocated by the BS. Its transmission also includes the buffer status of the device. We assume that all packets have the same size of $L$ bits.  
After a guard interval, the BS acknowledges the reception of successful transmissions.  

The set of active users at frame $t\in\mathbb{N}$ is denoted $\mathcal{U}(t)$ and the number of active devices is denoted $U(t)$, i.e., $|\mathcal{U}(t)|=U(t)$. We denote also $\boldsymbol{u}(t) \in \{0,1\}^K$ the vector of active users at frame $t$ such that $u_{k}(t) = \mathds{1}\{k \in \mathcal{U}(t)\}$. Besides, we define $\boldsymbol{\tau}^{p}(t)$, $\boldsymbol{\tau}^{a}(t)$, $\boldsymbol{\tau}^{s}(t)$ vectors of size $K$, where each component $k$ represents the number of frames since the last time device $k$ has been polled,  active and successfully decoded, respectively. 

We assume that the system is using NOMA~\cite{saito2013non} to improve the spectral efficiency of the network. NOMA allows several users to use the same frequency and time resources by superposing their signal in the power domain. At the receiver side, the BS applies Successive Interference Cancellation (SIC) to decode the superposed signals.

\subsection{Interference Channel Model}\label{section:interference}

We adopt a realistic channel model that has been adopted in the literature, based on the evaluation of the Signal-to-interference-plus-noise ratio (SINR)~\cite{salaun20} and the finite block length regime, see e.g.~\cite{ren2019joint}. 

\subsubsection{Received Signal}
In this model, a device $k\in \mathcal{U}(t)$, active in frame $t$, transmits a signal $s_k(t)$ of power $p_k(t)=\mathbb{E}[||s_k(t)||^2]$, where the expectation is taken over possible symbols. The BS is supposed to receive the signal with $n_a$ antennas and to perform Maximum Ratio Combining (MRC). The transmission of user $k$ experiences a large scale fading $g_k(t)$, which accounts for the distance-dependent path-loss and shadowing, fast fading $\boldsymbol{h}_k(t)=[h_{k1}(t),\cdots, h_{kn_a}(t)]^T\in \mathbb{C}^{n_a\times 1}$ and thermal noise $\boldsymbol{n}\in \mathbb{C}^{n_a\times 1}$. The signal received by the BS at frame $t$ from all active devices can thus be written as a superposition of $s_1(t), \dots, s_{U(t)}(t)$ and thermal noise: 
\begin{eqnarray}
\boldsymbol{r_s}(t) = \sum_{k\in\mathcal{U}(t)} \boldsymbol{h}_k(t)\sqrt{g_k(t)}s_k(t)+\boldsymbol{n}(t)
\end{eqnarray}
where $n_i(t), i=1,\dots, n_a$ is an independent circularly symmetric white Gaussian process with distribution $\mathcal{CN}(0,\sigma_n^2\boldsymbol{I})$. Thanks to the orthogonal pilots sent on the uplink, the BS is able to estimate the channel realizations of active devices. From now on, we assume that the BS has a perfect channel state information for decoding. In MRC, the signals received on the $n_a$ antennas are combined using a weight vector $\boldsymbol{w}_k^H=\boldsymbol{h}_k$ for device $k$. The combined signal $\boldsymbol{w}_k^H\boldsymbol{r_s}$ for device $k$ is thus:

\begin{equation} \label{eq:yk}
\boldsymbol{y}_k(t) = \boldsymbol{h}_k^H(t)\boldsymbol{h}_k(t)\sqrt{g_k(t)}s_k(t) + \boldsymbol{h}_k^H(t)\boldsymbol{n}(t) + \sum_{j\in\mathcal{U}(t)\backslash \{k\}} \boldsymbol{h}_k^H(t)\boldsymbol{h}_j(t)\sqrt{g_j(t)}s_j(t) 
\end{equation}

We assume that BS antennas are sufficiently spaced so that the fading coefficients at every antenna are spatially uncorrelated and thus $\boldsymbol{h}_k\sim\mathcal{CN}(0,\boldsymbol{I})$ for all $k$. The fast fading process $h_{ki}(t)$, for $k=1,...,K$ and $i=1,...,n_a$, is supposed to follow a time-correlated Gauss-Markov model~\cite{kobayashi2007joint}:
\begin{eqnarray}\label{eq:fast_fading_process}
h_{ki}(t) = \Bar{a}_kh_{ki}(t-1)+z_k(t)
\end{eqnarray}
where $z_k(t) \sim \mathcal{CN}(0,1 - \Bar{a}^2_k)$. The fading correlation coefficient $\Bar{a}_k$ is modeled using the Jakes' model~\cite{jakes94}: $\Bar{a}_k = J_0(2\pi v_k f_c T_f / c)$, where $J_0$ is the Bessel function of the first kind and order 0, $v_k$ is the speed of device $k$, $f_c$ is the carrier frequency, $c$ is the speed of light and $h_{ki}(0)\sim \mathcal{CN}(0,1)$. The coherence time for a device moving at speed $v$ is $T_c=c/(8f_cv)$~\cite{tse2005fundamentals}. The channels are supposed to be mutually independent across devices and constant during a frame (following a block fading channel model~\cite{goldsmith2005wireless}). We denote $\boldsymbol{H}(t)\in \mathbb{C}^{n_a\times K}$ the matrix of all channel realizations at time $t$ and $\mathcal{T}^H$ the evolution process, i.e., $\boldsymbol{H}(t+1)\sim\mathcal{T}^H(\boldsymbol{H}(t))$. 

\subsubsection{Decoding Order} \label{sec:decoding}
The SIC decoding order at each frame $t$ can be seen as a permutation function $\alpha(t)$ over the set of  active devices, i.e., $\alpha(t):[1:U(t)]\rightarrow \mathcal{U}(t)$. For any $i=1,...,U(t)$, $\alpha_i(t)$ is the $i$-th decoded device's index and for any $k\in\mathcal{U}(t)$, $\alpha_k^{-1}(t)$ is the rank of user $k$ in the decoding process. When the BS tries to decode device $\alpha_i(t)$, it has already tried to decode all devices $\alpha_1(t)$,..., $\alpha_{i-1}(t)$. Each decoding might have been successful or not. Let $\phi_k(t)$ be the indicator whether an active device $k\in\mathcal{U}(t)$ has been successfully decoded by the BS ($\phi_k(t)=1$) or not ($\phi_k(t)=0$) and $\boldsymbol{\phi}(t) = (\phi_1(t), \dots, \phi_K(t))$ the vector of all indicators. 
As a consequence, the signal received at the BS from $\alpha_i(t)$ is subject to the interference of $\alpha_j(t)$, $j>i$, i.e., from devices that have not been yet considered for decoding by the BS, and to the interference of $\alpha_j(t)$, $j<i$ whenever $\phi_{\alpha_j(t)}=0$, i.e., from devices that have not been successfully decoded by the BS.

We now assume the decoding order that minimizes the total transmit power, given target rates on the uplink~\cite{tse2005fundamentals}: active devices are sorted in decreasing order of their received power at the BS, as follows:
\begin{equation}
    \eta_{\alpha_1}(t)  \geq \eta_{\alpha_2}(t) \geq \dots \geq \eta_{\alpha_{U(t)}}(t)
\end{equation}
where $\eta_k(t)=p_k(t) g_k(t) ||\boldsymbol{h_k}(t)||^2$. We denote $\boldsymbol{\eta}(t) = (\eta_1(t), \eta_2(t), \dots, \eta_K(t))$ the vector of received powers. We denote $\boldsymbol{\eta}^o(t)$ the vector of powers received by the BS from the active devices, observed at time $t$  thanks to the transmitted pilots, i.e., $\boldsymbol{\eta^{o}}(t) = \text{diag}(\boldsymbol{u}(t)) \boldsymbol{\eta}(t)$.

\subsubsection{Signal to Interference plus Noise Ratio}

In absence of SIC, the signal \eqref{eq:yk} results in a SINR at the output of the combiner~\cite{tokgoz06}:
\begin{eqnarray}
\gamma_k^{\text{no-SIC}}(t) &=& \frac{\eta_k(t)}{\sum_{j\neq k}\eta_{jk}(t)+\sigma_n^2}
\end{eqnarray}
where $\eta_{jk}(t)=p_j(t)g_j(t)\frac{|\boldsymbol{h}_k^H\boldsymbol{h}_j|^2}{||\boldsymbol{h}_k||^2}$.

With SIC however, we decode in the decreasing order of $\eta_k(t)$, so that part of the interference is potentially successively removed. The SINR with SIC writes now:
\begin{eqnarray} \label{eq:sinr}
\gamma_k(t) = \frac{\eta_k(t)}{\underbrace{\sum_{j\in J_1} (1-\phi_j(t)) \eta_{jk}(t)}_{\text{before $k$  in decoding order}} + \underbrace{\sum_{j\in J_2}\eta_{jk}(t)}_{\text{after $k$}}+ \sigma_n^2}
\end{eqnarray}
where $J_1=\{j\in \mathcal{U}(t),\alpha_j^{-1}(t)<\alpha_k^{-1}(t)\}$  is the set of devices that are considered for decoding before $k$ and $J_2=\{j\in \mathcal{U}(t),\alpha_j^{-1}(t)>\alpha_k^{-1}(t)\}$ is the set of devices that are decoded after $k$. Note that $\phi_j$ is determined iteratively: we are able to compute the SINR of device $k$, only once we know the outcome of the decoding for devices $j \in J_1$.

\subsubsection{Achievable Rate with Finite Block Length}
As the URLLC messages are often supposed to be very small~\cite{3gpp.38.824} (in the factory automation scenario for instance), we adopt a finite block-length regime~\cite{polyanskiy2010channel} for the calculation of the achievable rate. In this model, an encoder maps every $L$-bit message $m\in [1:M]$ to a codeword $c_m\in \mathbb{C}^n$, where $M=2^L$ is the size of the message space and $n$ is the block length, also known as the number of complex channel uses. Codewords are subject to an average power constraint, i.e., $\frac{1}{M}\sum_m ||c_m||^2=n\eta$, where $\eta$ is the received power per channel use. The codeword is transmitted over an Average White Gaussian Noise channel with noise variance $\sigma^2$. At the receiver, a decoder maps the channel output to an estimate $\tilde{m}$ of the message. The average error probability is defined as $\varepsilon=\mathbb{P}[\tilde{m}\neq m]$. A codebook $\{c_m\in \mathbb{C}^n, m\in [1:M]\}$ and a decoder whose average error probability is less than $\varepsilon$ are called a $(M,n,\varepsilon)$-code. For given $\varepsilon$ and $n$, the maximum code size is denoted $M^*(n,\varepsilon)$. 
Authors of \cite{polyanskiy2010channel} provide a normal approximation of the maximum achievable code rate:
\begin{eqnarray} \label{eq:maxrate}
\frac{\log_2 M^*(n,\varepsilon)}{n}&\approx & C(\gamma)-\sqrt{\frac{V(\gamma)}{n}}Q^{-1}(\varepsilon)
\end{eqnarray}
where $\gamma=\frac{P}{\sigma^2}$ is the Signal to Noise Ratio (SNR), $P$ is the received power, $C(\gamma)=\log_2(1+\gamma)$ is the Shannon capacity\footnote{Note that contrary to \cite{polyanskiy2010channel}, which considers the capacity per real dimension, we use a complex representation of the signal. As a consequence, $C$ is the capacity per complex dimension~\cite[Chapter 5]{tse2005fundamentals}.}, $V(\gamma)=\frac{\gamma}{2}\frac{\gamma+2}{(\gamma+1)^2}\log_2^2e$ is the channel dispersion and $Q(x)=1/\sqrt{2\pi}\int_x^\infty \exp(-t^2/2)dt$. Although \eqref{eq:maxrate} is an asymptotic approximation when $n$ tends to infinity, it is tight for $n$ as small as $200$~\cite{polyanskiy2010channel}. When considering a block fading channel with channel realization $h$, \eqref{eq:maxrate} is valid \emph{conditionnally} to the channel realization with $\gamma=\frac{P|h|^2}{\sigma^2}$.  In our study, we further treat interference as noise, as it is usually done in the literature~\cite[Chapter 15]{goldsmith2005wireless}, and apply \eqref{eq:maxrate} with the SINR in \eqref{eq:sinr}.
As a consequence, a packet of device $k$, transmitted at frame $t$, is not successfully decoded with probability:
\begin{eqnarray} \label{eq:epsilon}
\epsilon_k(t)&=& Q\left(\sqrt{\frac{n}{V(\gamma_k(t))}}\left(C(\gamma_k(t))-\frac{L}{n}\right)  \right)
\end{eqnarray}
The downlink allocation and the acknowledgment are supposed to be error free.

\subsubsection{SIC Limitation}
 We define an upper limit $B$ on the number of possible multiplexed users which is characteristic of the SIC performance~\cite{tse2005fundamentals}. More specifically, a necessary condition for a device $k$ to be decoded is that the number of active devices is less than $B$, i.e.
 \begin{equation}\label{sic_limitation}
 |\mathcal{U}(t)| \leq B  
 \end{equation}

Finally, at frame $t$ and for a user $k$, we can write $\phi_k(t)$ as a Bernoulli random variable of parameter $\epsilon_k(t)$, i.e., $\phi_k(t)\sim \mathcal{B}(1-\epsilon_k(t))$ when~\eqref{sic_limitation} is satisfied, and $\phi_k(t)=0$ otherwise.

\subsection{Traffic Models}\label{section:traffic}

Packets are generated at the devices according to models of either probabilistic periodic traffic or probabilistic aperiodic traffic and are subject to a strict deadline constraint.

\subsubsection*{Probabilistic periodic traffic}
In this model, directly inspired by~\cite{hou2013packets}, a device $k$ generates packets periodically every $N_p$ radio frames with probability $q_k$. Devices are not synchronous, i.e., each device is assigned an offset parameter $f_k \in [0, N_p]$ such that, at every radio frame $t \geq 0$, the probability for a device $k$ of generating a new packet is: $\bar{q}_k(t | f_k, q_k, N_p) = \mathds{1}_{\{t[N_p] = f_k\}} q_k$. Note that a specific case for this model is the \emph{deterministic periodic traffic} as defined in \cite[Annex A]{3gpp.38.824} for various use cases including for example factory automation, where $q_k=1$ and $f_k=0$ for all $k$. Periodic transmissions may correspond to the periodic update of a position or the repeated monitoring of a characteristic parameter~\cite{3gpp.22.804}. 

\subsubsection*{Probabilistic aperiodic traffic}
This traffic model is defined in \cite{3gpp.38.824} and is based on the File Transfer Protocol (FTP) model 3 defined in \cite{3gpp.36.889}, however with a fixed packet length. At every device $k$, packets are generated according to a Poisson process of rate $\lambda_k$. An aperiodic transmission may correspond to process, diagnostic or maintenance events that trigger the transmission~\cite{3gpp.22.804}. 

\subsubsection*{Deadlines}
Every device $k$ has an individual latency constraint $\delta_k\in\mathbb{N}^*$ expressed in number of radio frames, such that a packet that has not been transmitted after $\delta_k$ radio frames is dropped. Let $\delta = \max_k{\delta_k}$. When a transmission fails, a device is allowed to retransmit the packet as long as it has not expired.

\subsubsection*{Buffers} 
We assume that devices have an infinite buffer and packets in the queue are delivered in a ``first come, first served" manner. For every device $k$, the buffer at time $t$ can be represented by a vector $\boldsymbol{b}_k(t)=[b_{k,1}(t),...,b_{k,\delta}(t)]\in \mathbb{N}^\delta$ where $b_{k,d}(t)=i$ when device $k$ has $i$ packets with time-to-deadline $d$ at time $t$. We denote $\boldsymbol{B}(t)$ the matrix of all buffer status at time $t$. The {\it head-of-line delay} $d^{h}_k(t)$ of user $k$ at frame $t$ is defined as $b_{k,d^h_k(t)}(t)\neq 0$ and $b_{k,d}(t)=0$ for all $d<d^h_k(t)$. This is the smallest time-to-deadline in the buffer of device $k$. We note $\boldsymbol{d^h} = [d^h_1, d^h_2, \dots, d^h_K]$ the vector of all head-of-line delays. When a device is polled, it chooses for transmission one of the packets associated to its head-of-line delay at random.

For a device $k$, the buffer status transits as follows: (a) Successfully decoded packets are removed from the buffer, i.e. $b_{k,d^h_k(t)-1}(t+1)=b_{k,d^h_k(t)}(t)-1$ if $\phi_k(t)=1$; (b) Other packets see their time-to-deadline decreased by one, i.e., $b_{k,d-1}(t+1)=b_{k,d}(t)$ for all $d>1$. If $d=1$, the packets expire and are removed from the buffers; (c) If $m$ new packets are generated at the device, they enter the buffer with a deadline $\delta_k$, i.e., $b_{k,\delta_k}(t+1)=m$. 

We denote this operation $\mathcal{T}^B$, i.e., 
\begin{equation}
    \boldsymbol{B}(t+1) \sim \mathcal{T}^B (\boldsymbol{B}(t), \boldsymbol{\phi}(t))
\end{equation}

When an active device is successfully decoded, its buffer status is known (or observed) to the BS. We thus denote $\boldsymbol{B}^o(t)$ the matrix of observed buffer status at time $t$.

\section{Problem Formulation}\label{section:problem_formulation}

\subsection{Optimization Problem}

The objective of the BS is to maximize the expected number of successful transmissions with respect to the stochastic policy $\pi$ that maps the current observation history at frame $t$: $(\boldsymbol{B}^o(t), \boldsymbol{\eta}^o(t), \dots, \boldsymbol{B}^o(0), \boldsymbol{\eta}^o(0))$ to the vector of devices to schedule $\boldsymbol{a}(t)$. Moreover, buffers and channels are subject to the dynamics $\mathcal{T}^B$ and $\mathcal{T}^H$, respectively. The optimization problem (\ref{P}) can thus be formulated as:
\begin{equation}
\begin{aligned}
\max_{\pi} \quad &\underset{(\mathcal{T}^B, \mathcal{T}^H, \pi)}{\mathbb{E}} \left[ \sum_{t=0}^\infty \sum_{k\in\mathcal{U}(t)} \gamma^t\phi_k(t)\right]\\
\text{s.t. } 
& \boldsymbol{B}(t+1) \sim \mathcal{T}^B(\boldsymbol{B}(t), \boldsymbol{\phi}(t)) \\
& \boldsymbol{H}(t+1) \sim \mathcal{T}^H(\boldsymbol{H}(t))\\
\end{aligned}
\tag{P}\label{P}
\end{equation}
where $\gamma\in[0,1)$ is the discount factor that determines the importance of future rewards compared to immediate ones.

\subsection{POMDP Formulation}
To solve our problem, we adopt the POMDP framework, see e.g.~\cite{sondik1971optimal, kaelbling1998planning}: at every time step $t$, an agent interacts with the environment by taking an action, gets an observation from the environment and obtains a reward.  

\begin{definition}[POMDP]
    A POMDP can be described by a tuple ($\mathcal{S}$, $\mathcal{A}$, $\mathcal{T}$, $\mathcal{R}$, $\Omega$, $\mathcal{O}$), where 
    \begin{itemize}
        \item $\mathcal{S}$ is the state space, i.e., a finite set of environmental states,
        \item $\mathcal{A}$ is the action space, i.e., a finite set of actions,
        \item $\mathcal{T}: \mathcal{S}\times \mathcal{A} \mapsto \Delta(\mathcal{S})$ is the transition function which is a probability distribution over the next environmental state $s'$ when it was in state $s$ and the action $a$ has been taken. It verifies the Markov property: $ s' \sim \mathcal{T}(\cdot | s(t)=s,a(t)=a)$, 
        \item $\mathcal{R}: \mathcal{S} \times \mathcal{A} \mapsto \mathbb{R}$ is the reward function, where $\mathcal{R}(s,a)$ is the immediate reward by taking action $a$ in state $s$. We denote $r(t)$ the immediate reward at time $t$.
        \item $\Omega$ is the observation space, i.e., a finite set of observations,
        \item $\mathcal{O}: \mathcal{S}\times \mathcal{A} \mapsto \Delta(\Omega)$ is the probability distribution of the observation $o$ when the environment is in state $s'$ and the agent has taken action $a$: $o \sim \mathcal{O}(\cdot |s(t+1)=s',a(t)=a)$.
    \end{itemize}
\end{definition}

The \emph{history} $\hbar(t)$ at time $t$ is defined as the sequence of actions taken by the agent and observations from the environment $\hbar(t)=\{o(0),a(0), o(1), a(1),...,a(t-1),o(t)\}$, where $a(t)\in\mathcal{A}$ and $o(t)\in\Omega$ for all $t$. The agent makes decisions using a stochastic \textit{policy} $\pi$ that is a distribution over the actions knowing the history. 

The POMDP related to our problem consists in the following components. 

\subsubsection{State space} At each step $t$, a state $\boldsymbol{s}(t)$ is defined as the concatenation of the buffer status $\boldsymbol{B}(t)$, the received powers $\boldsymbol{\eta}(t)$, and the observation $\boldsymbol{o}(t)$ obtained from the active users at the previous step: 
\begin{equation}
    \boldsymbol{s}(t) = \langle \boldsymbol{B}(t), \boldsymbol{\eta}(t), \boldsymbol{o}(t) \rangle
\end{equation}
\noindent where $\boldsymbol{o}(t) = \langle \boldsymbol{u}(t-1), \boldsymbol{\phi}(t-1), \boldsymbol{B}^{o}(t-1), \boldsymbol{\eta}^{o}(t-1), r(t-1) \rangle$ is the vector of active users, the vector of decoded packets, the observed buffers, the observed received power and the reward at $t-1$ respectively. 

\subsubsection{Action space} The agent has the possibility to poll any subset of devices at every frame. The action space is thus defined as $\mathcal{A}=\{0, 1\}^K$. For $\boldsymbol{a}=(a_1, a_2, \dots, a_K) \in \mathcal{A}$, $a_k=1$ if the agent polls device $k$ and $a_k=0$ otherwise. Note that the action space grows exponentially with the number of devices.

\subsubsection{Transition function} When the system is in state $\boldsymbol{s}(t)$ at the beginning of a radio frame $t$, it transits to state $\boldsymbol{s}(t+1)$ at the end of the radio frame. The received power $\boldsymbol{\eta}(t)$ evolves with the channel realizations $\boldsymbol{H}(t)$ and is governed by $\mathcal{T}^H$. Finally the evolution of the buffers is described by $\mathcal{T}^B$. The next observation  $\boldsymbol{o}(t+1)$ is computed using the observation function defined in the next subsection.

\subsubsection{Observation space and observation function}
At every frame $t$, the RL agent can only observe the last feedback from the active users: the set of active users, their channel realizations, the buffer status of successfully decoded devices and the reward. From the state $\boldsymbol{s}(t+1)$ and action $\boldsymbol{a}(t)$, the observation at time $t+1$ is deterministic and defined by: 
\begin{align}
    \boldsymbol{o}(t+1) &= \mathcal{O}(\boldsymbol{s}(t+1),\boldsymbol{a}(t)) \\
    &= \langle \boldsymbol{u}(t), \boldsymbol{\phi}(t), \boldsymbol{B}^{o}(t), \boldsymbol{\eta}^{o}(t), r(t) \rangle
\end{align}

In particular, $\boldsymbol{\phi}(t)\sim\mathcal{B}(1 - \boldsymbol{\epsilon}(t))$, $\boldsymbol{B^{o}}(t) = \text{diag}(\boldsymbol{u}(t)\odot \boldsymbol{\phi}(t))\boldsymbol{B}(t)$ and $\boldsymbol{\eta^{o}}(t) = \text{diag}(\boldsymbol{u}(t)) \boldsymbol{\eta}(t)$.
Note that when a device $k$ is active but its packet is not decoded, the agent still has the information that this device has a packet to transmit through $u_k(t)=1$.

\subsubsection{Reward function} 
We define the reward function as the number of successfully decoded packets:
\begin{equation}
    \mathcal{R}(\boldsymbol{s}(t), \boldsymbol{a}(t)) = \sum_{k\in\mathcal{U}(t)}\phi_k(t) %- \frac{1}{K} \sum_{k=1}^K \mathds{1}\{\tau^{a}_{k}(t) > \delta_k\}
\end{equation}

Note that unlike most RL approaches for multiple access \cite{xu2020application, chang2018distributive, guo2022multi}, we do not penalize the agent when there is a collision or interference. The reason is that we want the agent to learn a tradeoff between sensing and transmitting. In our experiments, we have noticed that using a penalty for collisions did not improve the performance. 

The POMDP formulated above aims at maximizing the optimization problem (\ref{P}). 
In general, POMDP problems are known to be PSPACE-complete \cite{papadimitriou1987complexity}, which means that they can be solved using a polynomial amount of memory space and are at least as hard as every other PSPACE problem. 
In order to solve this POMDP, we introduce a sufficient statistic for the history of past actions and observations, that we call the \emph{agent state}, and that allows us to transform the POMDP problem into an MDP.

\subsection{Agent state for Solving a POMDP}

\begin{definition}[Agent state] \label{def:agentstate}
    At the beginning of each frame $t\geq 1$, we define the \emph{agent state} $\boldsymbol{A}(t)$ after the agent receives its observation $\boldsymbol{o}(t)$ as:
\begin{equation}
    \boldsymbol{A}(t) = \langle \boldsymbol{B}^A(t), \boldsymbol{\eta}^A(t), \boldsymbol{\tau}^{p}(t), \boldsymbol{\tau}^{a}(t), \boldsymbol{\tau}^{s}(t), r(t-1) \rangle,
\end{equation}
where $\boldsymbol{\tau}^{p}(t)$, $\boldsymbol{\tau}^{a}(t)$ and $ \boldsymbol{\tau}^{s}(t)$ are the number of frames from $t$ since the last time the devices have been polled, active and have successfully transmitted respectively. $\boldsymbol{\eta}^A(t)$ is the last known received power of the active devices, i.e., its $k$-th column is $\boldsymbol{\eta}_{k}(t-{\tau}^{a}_k(t)-1)$. The matrix $\boldsymbol{B}^A(t)$ is a representation of the buffers given the observations made by the BS at time $t$ and is defined as follows. If an active user $k$ has been successfully decoded in the previous frame, we update $\boldsymbol{b}^A_k(t)$ with the new observation, i.e., $\boldsymbol{b}^A_k(t) = \boldsymbol{b}^{o}_k(t-1)$. For all devices, we decrease the deadlines of the packets in the buffers representation at time $t-1$ by 1 and we remove the expired packets. 
\end{definition}
%For $t=0$, we set $\boldsymbol{A_0}=\boldsymbol{O_\mathbb{A}}$.

While $\boldsymbol{B}^o(t)$ is an immediate observation of the buffers of the active users at $t$, $\boldsymbol{B}^A(t)$ is a compact representation of all past buffer observations at $t$. Introducing $\boldsymbol{B}^A(t)$ allows us to incorporate the knowledge of the dynamics of the buffers in the agent. Yet, the agent is still not aware of the new arrivals. 
To summarise, the agent state at frame $t$, $\boldsymbol{A}(t)$, can be written as a function $f^A$ of the observation $\boldsymbol{o}(t)$, the previous action $\boldsymbol{a}(t-1)$ and the previous agent state $\boldsymbol{A}(t-1)$, i.e., 
\begin{equation} \label{eq:agentstateevolution}
    \boldsymbol{A}(t) = f^A(\boldsymbol{A}(t-1), \boldsymbol{o}(t), \boldsymbol{a}(t-1))
\end{equation}

\begin{proposition} \label{prop:sufficient}
    $\boldsymbol{A}$ is a sufficient statistic for the action-observation history, i.e.,
    \begin{equation}
        P(s(t) | \hbar(t)) = P(s(t) | A(t))
    \end{equation}
\end{proposition}

\begin{IEEEproof}
    See Appendix~\ref{app:sufficient}. 
\end{IEEEproof}

\begin{proposition}
    The tuple $(\mathcal{S}^A, \mathcal{A}, \mathcal{T}^A, \mathcal{R}^A)$ forms an MDP where $\mathcal{T}^A: \mathcal{S}^A \times \mathcal{A} \mapsto \Delta(\mathcal{S}^A)$ is the agent state transition function and $\mathcal{R}^A: \mathcal{S}^A \times \mathcal{A} \mapsto \mathbb{R}$ the agent state reward function such that:
    \begin{align*} 
        \mathcal{T}^A(\boldsymbol{A}(t), \boldsymbol{a}(t)) = \sum_{\boldsymbol{o}(t+1) \in \Omega} f^A(\boldsymbol{A}(t),  \boldsymbol{a}(t), \boldsymbol{o}(t+1)) P(\boldsymbol{o}(t+1) | \boldsymbol{a}(t), \boldsymbol{A}(t))
    \end{align*}  
    \begin{align*}
        \mathcal{R}^A(\boldsymbol{A}(t), \boldsymbol{a}(t)) = \sum_{\boldsymbol{s}(t) \in \mathcal{S}}P(\boldsymbol{s}(t) | \boldsymbol{A}(t)) \mathcal{R}(\boldsymbol{s}(t), \boldsymbol{a}(t))
    \end{align*}

\noindent where:
\begin{align*}
P(\boldsymbol{o}(t+1) | \boldsymbol{a}(t), \boldsymbol{A}(t)) =\
\sum_{\boldsymbol{s} \in \mathcal{S}} \mathcal{O}(\boldsymbol{o}(t+1) | \boldsymbol{s}, \boldsymbol{a}(t)) \sum_{\boldsymbol{s} \in \mathcal{S}} \mathcal{T}(\boldsymbol{s}(t+1) | \boldsymbol{s}, \boldsymbol{a}(t))
\end{align*}

\end{proposition}

\begin{IEEEproof}
    The expressions of $\mathcal{T}^A$ and $\mathcal{R}^A$ are directly derived using the law of total probability and the Bayes formula.
\end{IEEEproof}

\begin{figure}[t]
    \centering
\includegraphics[width=0.9\linewidth]{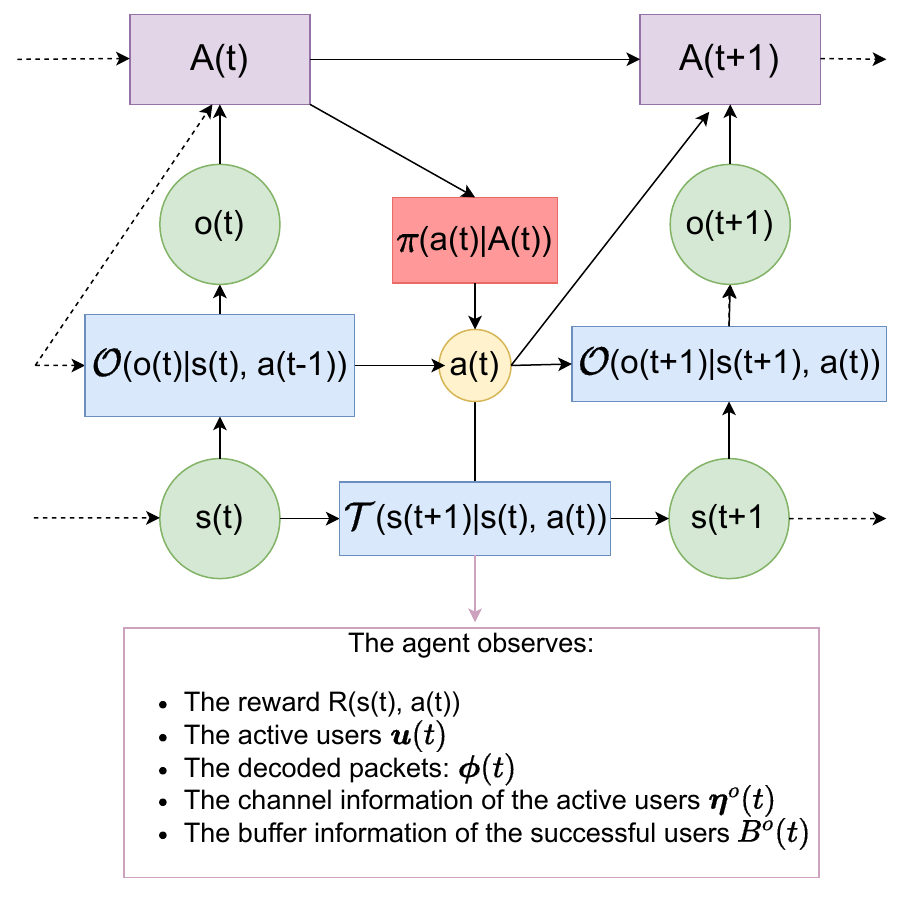}
    \caption{Formulation of the NOMA-URLLC problem.}
    \label{fig:pomdp}
\end{figure}

The problem formulation is summarized in Fig.~\ref{fig:pomdp}. 
\begin{enumerate}
    \item At the beginning of frame $t$, the system is at state $\boldsymbol{s}(t)$.
    \item The agent can observe $\boldsymbol{o}(t) = \mathcal{O}(\boldsymbol{o}(t) | \boldsymbol{s}(t), \boldsymbol{a}(t-1))$.
    \item It then computes the agent state using \eqref{eq:agentstateevolution}. %$\boldsymbol{A}(t) = f^A(\boldsymbol{A}(t-1), \boldsymbol{o}(t), \boldsymbol{a}(t-1))$.
    \item It makes an action $a(t) \sim \pi(\boldsymbol{a}(t) |\boldsymbol{A}(t))$.
    \item The system then transitions to the next state as follows: $\boldsymbol{s}(t+1) \sim \mathcal{T}(\boldsymbol{s}(t+1) |\boldsymbol{s}(t), \boldsymbol{a}(t))$ and the agent receives: $R(\boldsymbol{s}(t), \boldsymbol{a}(t)), \boldsymbol{u}(t), \boldsymbol{\phi}(t), \boldsymbol{\eta}^o(t), \boldsymbol{B}^o(t)$.
\end{enumerate}

Transforming the POMDP problem in an MDP allows us to leverage DRL algorithms in order to solve the optimization problem~(\ref{P}).

\section{Deep Reinforcement Learning Approach}\label{section:nomappo}

\subsection{Proximal Policy Optimization algorithm}

The Proximal Policy Optimization (PPO) algorithm \cite{schulman2017proximal} is a Policy Gradient (PG) algorithm that benefits from the Trust Region Policy Optimization policy update \cite{schulman2015trust} while being data efficient. The idea is to restrict the amplitude of the policy update in order to improve training stability while only using first-order optimization. PPO maximizes the following objective with respect to parameters $\theta$: 
\begin{equation}\label{eq:ppo_obj}
     \mathbb{E}_{\boldsymbol{s},\boldsymbol{a} \sim (\pi_{\theta_\text{old}}, \mathcal{T})}\!\! \left[ \min \! \left(\frac{\pi_{\theta}(\boldsymbol{a}|\boldsymbol{s})}{\pi_{\text{old}}(\boldsymbol{a}|\boldsymbol{s})} \!A^{\pi_{\theta_\text{old}}}(\boldsymbol{s}, \boldsymbol{a}), g(\nu) A^{\pi_{\theta_\text{old}}}(\boldsymbol{s}, \boldsymbol{a})\! \right)\! \right]
\end{equation}

\noindent with $g(\nu) = \text{clip} \left(\frac{\pi_{\theta}(\boldsymbol{a}|\boldsymbol{s})}{\pi_{\text{old}}(\boldsymbol{a}|\boldsymbol{s})}, 1-\nu, 1+\nu \right)$ and $\nu \in [0,1)$ a hyperparameter that indicates how far away the new policy can deviate from the old one.
$A^{\pi_{\text{old}}}$ is the advantage function and is defined by: 
\begin{eqnarray} \label{eq:advfunc}
A^{\pi_{\text{old}}}(\boldsymbol{s}(t), \boldsymbol{a}(t)) = Q^{\pi_{\text{old}}}(\boldsymbol{s}(t), \boldsymbol{a}(t)) - V^{\pi_{\text{old}}}(\boldsymbol{s}(t))
\end{eqnarray}

It describes the value $Q^{\pi_{\theta_\text{old}}}(\boldsymbol{s}(t), \boldsymbol{a}(t))$ of an action $\boldsymbol{a}$ in a state $\boldsymbol{s}$  compared to the value of the state $V^{\pi_{\theta_\text{old}}}(\boldsymbol{s}(t))$ (how much better or worse it is to take this action). The estimated advantage function, noted $\hat{A}$ can be computed according to several methods that can be found in \cite{schulman2015high}. The most efficient method, which is also the one we adopt in NOMA-PPO, is the Generalized Advantage Estimation (GAE) algorithm \cite{schulman2015high}. This algorithm uses the temporal difference residuals $\delta^V(t) = r(t) - \gamma V(\boldsymbol{s}(t+1)) - V(\boldsymbol{s}(t))$ in order to define the \emph{Generalized Advantage Estimator} $\hat{A}^{GAE}(t)$:
\begin{equation}
    \hat{A}^{GAE}(t) = \sum_{l=0}^\infty (\gamma\lambda_{GAE})^l \delta^V(t+l)
\end{equation}

\noindent where $\lambda_{GAE} \in [0, 1]$ adjusts the bias-variance tradeoff. This method manages to reduce the variance of the gradient estimate and stabilizes training at the cost of introducing a bias. In practice, the value function $V$ is approximated by a DNN with parameters $\varphi$: $V_{\varphi}$.

\subsection{Exploiting Prior Knowledge}\label{section:prior}

In our scheduling problem, the Earliest Deadline First (EDF) scheduler, which schedules pending packets in the increasing order of their deadline,  
intuitively is a good heuristic when the environment is fully observable by the scheduler. EDF is indeed known to be optimal in various deterministic~\cite{stankovic1998deadline} and stochastic (see e.g.~\cite{moyal2013queues}) settings. 
We thus adapt it to NOMA as follows: given the devices' buffers, $\boldsymbol{B}(t)$,  EDF schedules the $B$ users with the smallest head-of-line delay $d_k^h(t)$.
\begin{align} 
\label{EDF}
    \operatorname{EDF}(\boldsymbol{B}(t)) &= (a_1, \dots, a_K),\\
    \text{where } a_k &= \left\{
    \begin{array}{ll}
        1 & \mbox{if }k\in\arg_B \min(\{d_1^h(t), \dots, d_K^h(t)\}) \\
        0 & \mbox{otherwise}
    \end{array}
\right. \notag
\end{align}

Note that in our POMDP problem, EDF cannot be implemented in practice, as it requires full observability of the system, but can serve as a valuable benchmark.
We can further allow the scheduler to take into account the channel state, by introducing a prior regarding the channel quality. In particular, we define a prior on the channel $f_{\text{ch}}$ as follows:

\begin{align} 
\label{f_channel}
    f_{\operatorname{{\text{ch}}}}(\boldsymbol{\eta}(t), \boldsymbol{\tau^a}) &= (a_1, \dots, a_K),\\
    \text{where } a_k &= \left\{
    \begin{array}{ll}
        0 & \mbox{if }\eta_k \leq \eta^* \mbox{ and } \tau^a_k \leq \tau^* \\
        1 & \mbox{otherwise}
    \end{array}
\right. \notag
\end{align}

\noindent where $\eta^* \geq 0$ and $\tau^* \geq 0$ are hyperparameters to determine the quality of a channel. Typically, $\eta^*$ is the threshold that indicates when a user will not be decoded with a high probability, regardless of the others' channels and $\tau^*$ is the coherence time that indicates whether the last information we have on the channel is relevant or outdated. The intuition behind this prior is that a user should remain inactive if it experiences a very ``bad" channel. 

The resulting prior $f$ is thus a combination of the EDF and channel prior:

\begin{equation}
    \label{prior_f}
    f(\boldsymbol{a}; \boldsymbol{A}) = EDF(\boldsymbol{B}(t)) \odot f_{\operatorname{{\text{ch}}}}(\boldsymbol{\eta}(t), \boldsymbol{\tau^a})
\end{equation}

In order to incorporate this prior knowledge into the RL agent, we introduce a Bayesian policy inspired by \cite{titsias2018bayesian}. We express the posterior policy $q(\boldsymbol{a}|\boldsymbol{A}; \theta_\pi)$ as a function of the prior over the agent state $f(\boldsymbol{a}; \boldsymbol{A})$ and the task specific policy $\pi(\boldsymbol{a}|\boldsymbol{A}; \theta_\pi)$ parameterized by $\theta_\pi$ with the Bayes rule: 
\begin{equation}
    q(\boldsymbol{a}|\boldsymbol{A}; \theta_\pi) \propto \pi(\boldsymbol{a} | \boldsymbol{A}; \theta_\pi) \odot f(\boldsymbol{a}; \boldsymbol{A}) 
\end{equation}

\subsection{Algorithm Overview and Architecture}

\begin{figure}[H]
    \centering
    \includegraphics[scale=0.7]{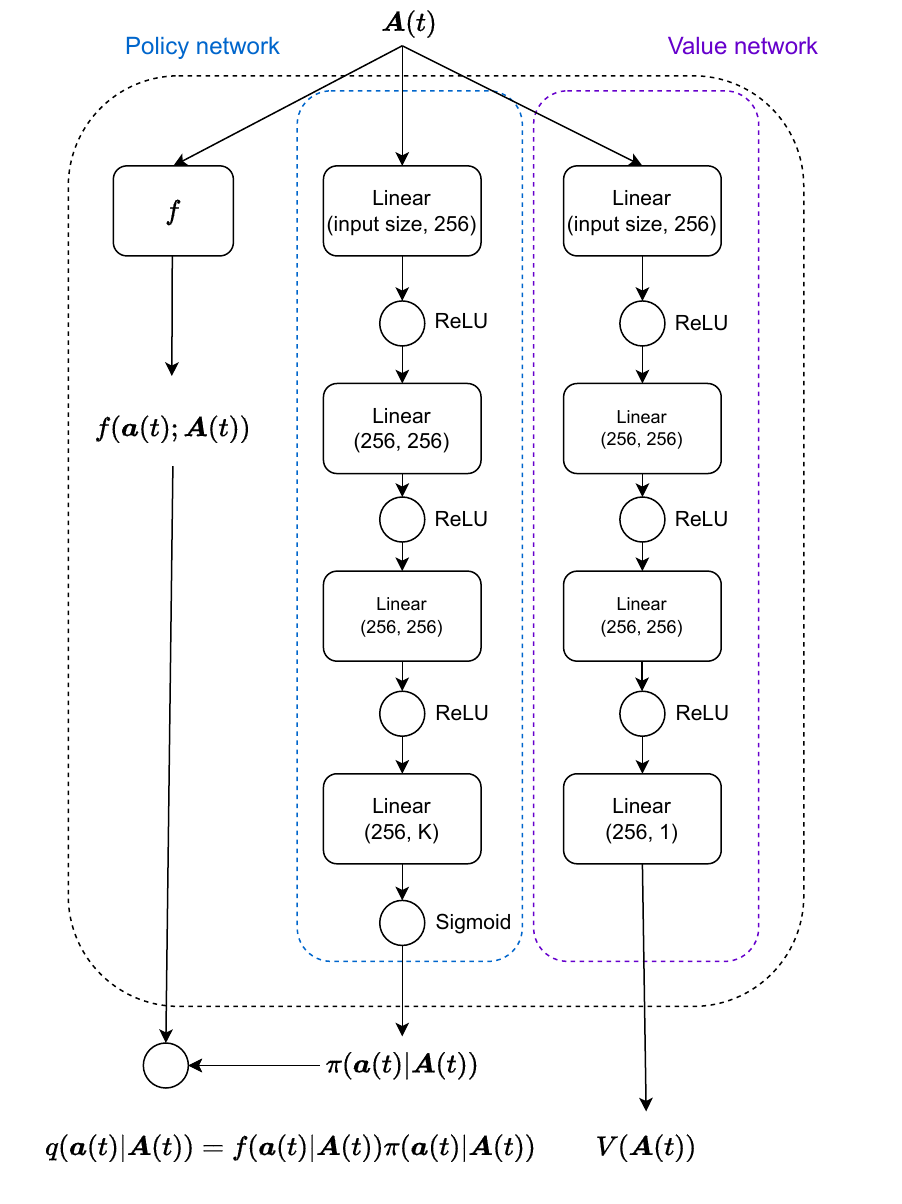}
    \caption{Architecture of the NOMA-PPO agent.}
    \label{fig:noma-ppo}
\end{figure}

The neural network architecture is described in Fig.~\ref{fig:noma-ppo}. NOMA-PPO uses two neural networks, one for the policy and one for the critic. The input vector is the concatenation of the preprocessed buffer information $\boldsymbol{B}^A(t)$, the timing information $1/\boldsymbol{\tau}^p(t)$,  $1/\boldsymbol{\tau}^a(t)$,  $1/\boldsymbol{\tau}^s(t)$, the channel information $\boldsymbol{\eta}^A(t)$ and the last reward $r(t-1)$. Its size is thus $5K+1$.

Following a branching architecture, the policy network produces activation probabilities for each user, yielding $K$ outputs: $\pi_\theta(\boldsymbol{a} |\boldsymbol{A}) = (\pi_\theta(a_1|\boldsymbol{A}), \pi_\theta(a_2 | \boldsymbol{A}) \dots, \pi_\theta(a_K | \boldsymbol{A}))$. Inspired by the BDQ architecture \cite{tavakoli2018action}, which employs this approach for Q-learning, we handle the combinatorial action space in a manner that scales linearly with the number of users and adapt it to the PPO algorithm. These $K$ outputs are then coordinated by a first block of hidden layers that are shared by all \textit{branches}. %\textcolor{purple}{Importantly, this adjustment allows for the definition of Bayesian policies and the integration of prior knowledge about the MA problem into the DRL agent, solutions that cannot be easily adapted to BDQ.}.

On the other hand, the value network follows the same architecture of the policy network, except that it outputs a single value for the state value estimation. The procedure for training NOMA-PPO is developed in Algorithm~\ref{algo-pseudocode}. The neural networks are updated $J$ times.  

\begin{algorithm}
\SetInd{0.25em}{0.15em}
\SetAlgoLined

\textbf{Input}: prior $f$, initial parameters of the policy network $\pi_{\theta_0}$ and the value network $V_{\varphi_0}$;\\

\For{$j = 1, 2, \dots, J$}{
    %\textcolor{magenta}{What is $J$? }
    Run the posterior policy $q_{\theta_j}$ and collect a set of $\beta$ trajectories $\{(\boldsymbol{A}_b(t), \pi_{\theta_j}(\boldsymbol{a}_b(t) | \boldsymbol{A}_b(t)), r_b(t))_{t=1,\dots, T}\}_{b=1\dots \beta}$.\\
    Compute the rewards-to-go $\hat{R}_b(t)$ for each trajectory: $\hat{R}_b(t) = \sum_{t'=t}^T \gamma^{t'} r_b(t')$\\
    Compute the values $V_{\phi_j}(\boldsymbol{A}_b(t))$ using the value network.\\
    Compute the advantage estimates $\hat{A}^{GAE}_b(t)$. \\
    Update the policy network by maximizing (\ref{eq:ppo_obj}) with the Adam algorithm \cite{kingma2014adam}:
\begin{align} 
\theta_{j+1} = \arg \max_\theta \frac{1}{\beta T} \Biggl[\sum_{b=1}^\beta \sum_{t=1}^T \min \Biggl(\frac{\pi_\theta(\boldsymbol{a}_b(t) | \boldsymbol{A}_b(t))}{\pi_{\theta_j}(\boldsymbol{a}_b(t) | \boldsymbol{A}_b(t))} \hat{A}^{GAE}_b(t),  g(\nu) \hat{A}^{GAE}_b(t)\Biggr)\Biggr]\notag
\end{align}

    }
    Update the value network by minimizing the mean-squared error with the Adam algorithm:
    \begin{equation}
       \varphi_{j+1} \! =\! \arg \min_\varphi \!\frac{1}{\beta T} \!\sum_{b=1}^\beta \sum_{t=1}^T \!\left(V_{\varphi}(\boldsymbol{A}_b(t)) \!-\! \hat{R}_b(t) \right)^2
    \end{equation}

\caption{NOMA-PPO for URLLC uplink scheduling in NOMA systems.}
\label{algo-pseudocode}
\end{algorithm}

\section{Experiments}\label{section:experiments}

\subsection{Simulation Settings and Implementation Details}

{\small \ctable[
cap = Simulation Settings.,
caption = Network Simulation Settings.,
label = tab:simparam,
pos = ht
]{|c|c|c|}{
\tnote[a]{For a channel bandwidth of 40~MHz, the signal occupies 38.16~MHz after having excluded guard bands~\cite{3gpp.38.104}.}
\tnote[b]{Typical delay spread for an indoor hot-spot scenario with carrier frequency 4 GHz~\cite{3gpp.38.824}.}
\tnote[c]{Normal cyclic prefix duration for symbols not at the start or in the middle of the subframe~\cite{3gpp.38.211}.}
\tnote[e]{Size of an array of size $56$ corresponding to a maximum deadline of $56$ frames, i.e., $10$~ms, with $3$~bits entries giving the number of packets for every deadline.}
\tnote[d]{Headers include 2~bytes of CRC~\cite{3gpp.38.212}, 1~byte for MAC~\cite{3gpp.38.321}, 0~byte for RLC~\cite{3gpp.38.322}, 2~bytes for PDCP~\cite{3gpp.38.323}, 0~byte for SDAP~\cite{3gpp.37.324} and 40~bytes for IPv6~\cite{ipv6}.}
}{
\hline
Parameter & Notation & Value \\
\hline
Carrier frequency & $f_c$ & $4$~GHz \\
Bandwidth\tmark[a] & $W$ & $38.16$ MHz \\
Subcarrier spacing &  $\Delta f$ & $30$ kHz \\
Delay spread\tmark[b] & $T_d$ & $100$~ns \\
OFDM symbol information part & $T_i$ & $33.33$~$\mu$s \\ 
OFDM symbol cyclic prefix\tmark[c] &  $T_{cp}$ & $2.34$~$\mu$s \\
SIC limitation & $B$ & $3$ \\
Information length & $L_i$ & $32$~bytes \\
Headers\tmark[d] length & $L_h$ & $46$~bytes \\
Buffer information\tmark[e] & $L_b$ & $14$~bytes \\
Average inter-arrival rate & $1/\lambda$ & 2 ms\\ 
Deadline & $\delta$ & $1$~ms \\
Network layout & $\ell\times \ell'$ & $50 \times 120$~m$^2$ \\
Noise Power Spectral Density & $N_0$& $-174$~dBm/Hz\\
BS noise figure & $N_F$ & $5$~dB \\
BS antenna height & $\Tilde{h}_b$ & $3$~m \\
BS antenna gain & $G_b$ & $5$~dBi \\
BS number of antennas &   $n_a$ & 4 \\
Device transmit power & $p$ & $23$~dBm \\
Device antenna height & $\Tilde{h}_d$ & $1.5$~m \\
Device antenna gain & $G_d$ & $0$~dBi \\
Device speed & $v$ & $3$~km/h \\
\hline
}
}

\begin{table}[H]
    \caption{Parameters of the DRL algorithms.}
    \centering
\begin{tabular}{ |c|c| }
\hline
Parameter & Value \\
\hline
Input size ($H_{in}$) & $5 K +1$\\
Hidden size ($H$) & 256 \\
Discount factor ($\gamma$) & 0.3 \\
Learning rate actor & $10^{-4}$\\
Learning rate critic & $10^{-3}$\\
Batch size & 128 \\
History length & K\\
%Update target frequency & 100 \\
Episode length ($T$) & 200 slots \\
Training length ($J$) & 10k episodes\\
Activation functions & ReLU \\
Number of seeds & 5 \\
$\lambda_{GAE}$ & 0.95 \\
\hline

\end{tabular}
    \label{tab:parameters_rl}
\end{table}

Our simulation settings (see Table~\ref{tab:simparam}) adopt the parameters of the factory automation use case of the 3GPP 5G NR specifications on URLLC~\cite{3gpp.38.824} and industrial IoT~\cite{3gpp.38.825}. Our radio frame is made of five time-slots ($T_f=5T_s$), whose duration $T_s$ is equivalent to an OFDM symbol in NR. It can be decomposed into an information part of duration $T_i$ and a cyclic prefix of duration $T_{cp}$, which both depend on the subcarrier spacing $\Delta f$: $T_s=T_i+T_{cp}$ with $T_i=1/\Delta f$. From the signal bandwidth we substract the subcarriers dedicated to uplink pilots, so that, when there are $U$ polled devices and $n_p$ pilots per device, the number of complex channel uses is $n=(W-n_pU\Delta f)T_i$~\cite[Chapter 5]{tse2005fundamentals}. The number of pilots per device can be obtained as follows: $n_p=\lceil W/W_c \rceil$, where $W_c=1/(2T_d)$~\cite[Chapter 2]{tse2005fundamentals} is the coherence bandwidth and $T_d$ is the delay spread. 

Regarding the traffic model, we consider either a deterministic periodic traffic with period $1/\lambda$ or a probabilistic aperiodic traffic with average inter-arrival time $1/\lambda$. A packet can be decomposed into an information part of length $L_i$, a header part of length $L_h$, and a buffer description of length $L_b$, so that $L=L_i+L_h+L_b$. In URLLC, headers cannot indeed be neglected with respect to the message length. We assume that the information part, the header and the buffer information are jointly encoded~\cite{Popovski2019}. The traffic parameters of Table~\ref{tab:simparam} are taken from the factory automation use case of Release 16~\cite{3gpp.38.824}. 

For realistic numerical experiments, we partly adopt the scenario proposed in~\cite[Table A.2.2-1]{3gpp.38.824} for the factory automation use case, with a single BS. The network layout is a rectangle of size $\ell\times \ell'$; the BS is positioned at its center at a height $\Tilde{h}_b$ and serves devices, each at height $\Tilde{h}_d$ and moving with velocity $v$. Devices are uniformly distributed within the network area. Devices and BS benefit from antenna gains $G_{b}$ and $G_{d}$ respectively. For a speed of $v=3$~km/h, we obtain a coherence time of $T_c=c/(8f_cv)=11.2$~ms, which corresponds to $63$ radio frames. We choose an episode length of $200$ frames that allows us to consider speeds below $1$~km/h. The path-loss model is the ITU InH NLOS~\cite{3gpp.38.901}. The BS has a noise figure $N_F$, so that the noise power is $\sigma_n^2=N_0WN_F$, where $N_0$ is the noise power spectral density. Typical values for the channel parameters are given in Table~\ref{tab:simparam}. We express the deadlines and inter-arrival time in term of frames. Indeed, given the frame duration $T_f$, we can deduce that the average inter-arrival time of 2 ms corresponds to $11.2$ frames and that the deadline of $1$ ms to $5.6$ frames.

The parameters of the DRL algorithms are given in Table~\ref{tab:parameters_rl}. We preprocess the agent state as follows. In order to reduce the dimension of the buffer information, the matrix $\boldsymbol{B}^A(t)$ is transformed into a vector of size $K$ of head-of-line delays for each agent.  In order to improve stability of speed up training, we normalize $\boldsymbol{\tau}^p, \boldsymbol{\tau}^a, \boldsymbol{\tau}^s$ between 0 and 1 by taking $1/\boldsymbol{\tau}^p, 1/\boldsymbol{\tau}^a, 1/\boldsymbol{\tau}^s$. Finally, the channel threshold $\eta^*$ is calculated using (\ref{eq:epsilon}) such that the error probability in absence of interference corresponding to $\eta^*$ is equal to $10^{-5}$.

\subsubsection*{URLLC score}

In order to compare our algorithm to the traditional benchmarks, we define the \emph{URLLC score} as the number of successfully transmitted packets over the number of received packets. In the following experiments, the URLLC score is computed over 500 episodes which corresponds to approximately $2 \cdot 10^5$ generated packets according to the traffic parameters in Table~\ref{tab:simparam}. Therefore, a URLLC score of 1 means that the reliability is greater than $1-10^5$.

\subsection{Benchmarks}

For all baselines, when the BS receives two or more packets at the same time, we use the SIC procedure described in Section~\ref{section:interference} to decode the packets.

\begin{itemize}
    \item \textbf{Random Scheduler}: This scheduler schedules a subset of $B$ devices uniformly at random.

    \item \textbf{EDF}: This scheduler schedules pending packets in the increasing order of their deadlines, see \eqref{EDF}. Again, it cannot be implemented in practice on the uplink because of the assumed full observability of the device buffers.  
    
    \item \textbf{SA-NOMA-SIC}: This baseline is a grant-free approach that follows the work of \cite{tegos2020slotted}. It combines SA with SIC. At each frame, devices transmit their packet with the same probability $p$. Regarding re-transmissions, we use the \emph{proactive} scheme \cite{mahmood2019uplink}: a user can re-transmit the same packet with probability $p$ until it is delivered or expired. The probability $p$ is empirically optimized such that the URLLC score is maximized for every scenario.
    
    \item \textbf{RDQN-NOMA Scheduler}: The standard DQN algorithm proposed by \cite{hausknecht2015deep} is the traditional approach to solve POMDP problems. The idea is to use an RNN to handle partial observability. We directly apply this algorithm in order to solve (\ref{P}). The action space of the RL agent is the set of combinations of $B$ or more devices to poll.

    \item \textbf{Branching DQN (BDQ)}: this baseline is a version of the Dueling Double DQN algorithm from \cite{tavakoli2018action} that uses a branching architecture in order to handle a combinatorial action space. 
    
    \item \textbf{iDRQN-NOMA}: This baseline is a fully distributed Multi-Agent Reinforcement Learning (MARL) algorithm for grant-free multiple access that follows the solution of \cite{xu2020application} where each device is modeled by a Deep Q-network and decides to access the medium based on its local information: the state of its buffer and its channel state. This baseline uses a RNN, a Gated Recurrent Unit (GRU) layer \cite{chung2014empirical} in particular, as it is a standard approach to tackle partial observability. Additionally, we extend the work of \cite{xu2020application} to NOMA systems by adapting the reward function as follows: at the end of every frame $t$, every user $k$ receives the same reward:
    \begin{equation}
    R^k(s_k(t), a_k(t)) = \left\{
    \begin{array}{ll}
        \sum_{i\in\mathcal{U}(t)} \phi_i(t)  & \mbox{if } |\mathcal{U}(t)| \leq B\\ 
        -1 & \mbox{otherwise}
    \end{array}
\right.   
    \end{equation}

    \item \textbf{NOMA-PPO-no-prior}: This baseline is the proposed approach, however without using prior information over the agent state.
\end{itemize}

In order to be fair in the experiments, we modify the frame structure of the two grant-free approaches SA-NOMA-SIC and iDRQN-NOMA and divide it into four time-slots of duration $T_s$: an uplink transmission symbol, a guard symbol, a downlink ACK/NACK and a guard symbol. 

\subsection{Study of the Channel Model}

\begin{figure}[H]
\begin{subfigure}{.5\textwidth}
  \centering
  % include first image
  \includegraphics[width=1.\linewidth]{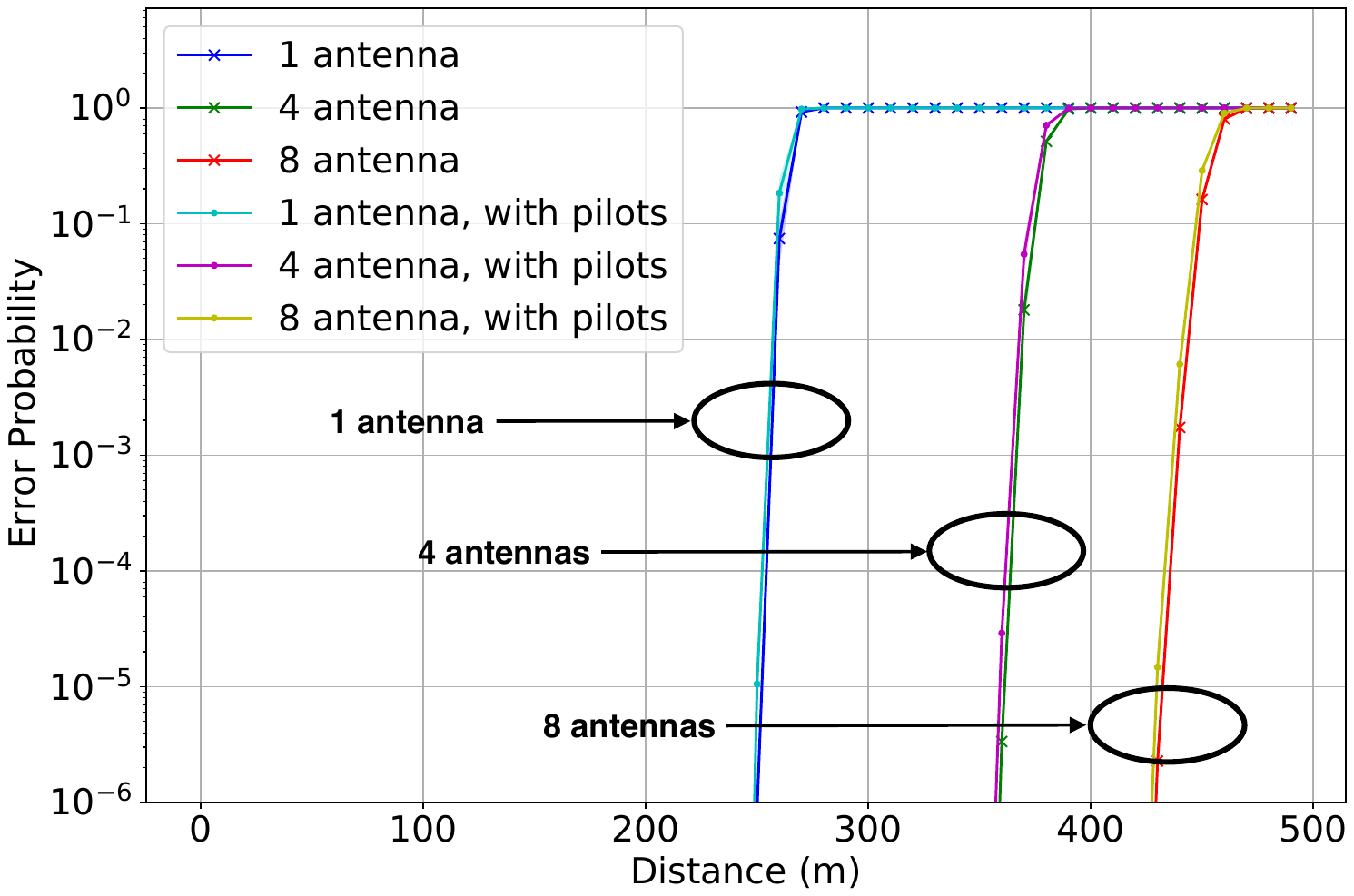}  
  \caption{For a single device.}
  \label{fig:channel_1_device}
\end{subfigure}
\begin{subfigure}{.5\textwidth}
  \centering
  % include second image
  \includegraphics[width=1.\linewidth]{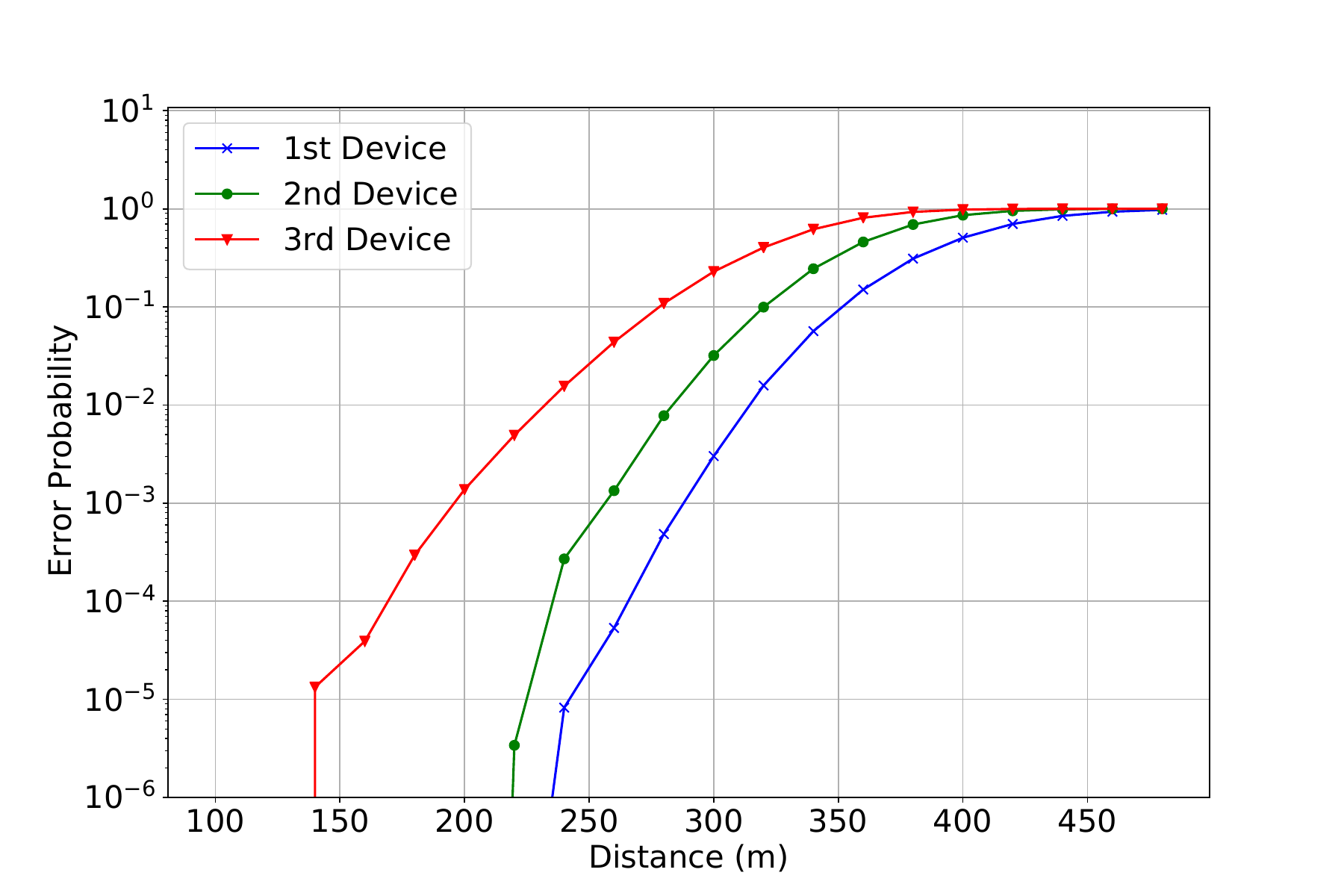}  
  \caption{For three devices after SIC.}
  \label{fig:channel_3_devices}
\end{subfigure}
\caption{Packet error probability $\epsilon$ as a function of the distance to the BS. }
\label{fig:channel_study}
\vspace*{-1cm}
\end{figure}

In this section, we study the behavior of the channel. In Fig.~\ref{fig:channel_1_device}, we show the channel error probability $\epsilon$ as a function of the distance between a device and the BS, involved in a point-to-point transmission without interference. Results are shown for different number of antennas at the BS and with or without the pilot signals. We see that there are roughly three regimes that can be distinguished. When the distance is small, the error probability is very small (less than $10^{-6}$). When the distance to the BS is too large, the error probability is close to $1$. In this regime, there is no hope to guarantee URLLC requirements. In an intermediate regime that depends on the number of antennas and the number of decoded devices, the error probability is not negligible but the URLLC requirements could be met with an appropriate scheduling. In this case, the SINR model is required to benefit from the channel evolution for every device. As expected, increasing the number of antennas at the BS improves the reliability. At last, reserving some resource for pilots has a negligible influence on the performance. 

In Fig.~\ref{fig:channel_3_devices}, we show the error probability as a function of the distance of the three devices from the BS. The three devices transmit simultaneously to the BS, which performs SIC. The resulting error probabilities are shown for the first, second and third decoded signals respectively. We again observe the three regimes, however with an offset according to the rank of decoding. The intermediate regime ranges here approximately between $150$~m and $300$~m.

\subsection{Convergence Analysis}

In Fig.~\ref{fig:training}, we show the evolution of the URLLC score during the training of 18 learning agents under the probabilistic aperiodic traffic. First, we can see that NOMA-PPO converges the fastest and with the smallest variance to its asymptotic value. Second, we see that not only does the prior help NOMA-PPO reach a better optimum, it also increases the convergence speed and reduces the variance. Third, while the MARL grant-free approach, iDRQN-NOMA, reaches the second best optimum in terms of URLLC score, it converges slower than NOMA-PPO and with a larger variance. This can be accounted for by the fact that agents must coordinate independently, solely using the BS's feedback. Furthermore, we observe that the DRQN-NOMA scheduler does not manage to converge due to the combinatorial action space. Indeed, there are $2^{K} - \sum_{k=0}^{B-1} \binom{K}{k} = 261,972$ possible actions for $K=18$ and $B=3$, thus choosing the appropriate action is challenging. In light of the lack of convergence of this algorithm, we exclude it from future experimental baselines. Finally, we observe that the BDQ algorithm comes third in term of asymptotic URLLC score but suffers from high variance. We notice that for the DRQN-NOMA scheduler and iDRQN-NOMA, the score does not evolve in the first thousand episodes. It is because of the ``warm up" stage where we collect trajectories in order to fill the replay buffer of the agents without updating them.

\subsection{Performance in the 3GPP Scenario}

\begin{figure*}[t]
    \centering
    \begin{subfigure}[b]{0.48\textwidth} 
    \centering
        \includegraphics[width=\textwidth]{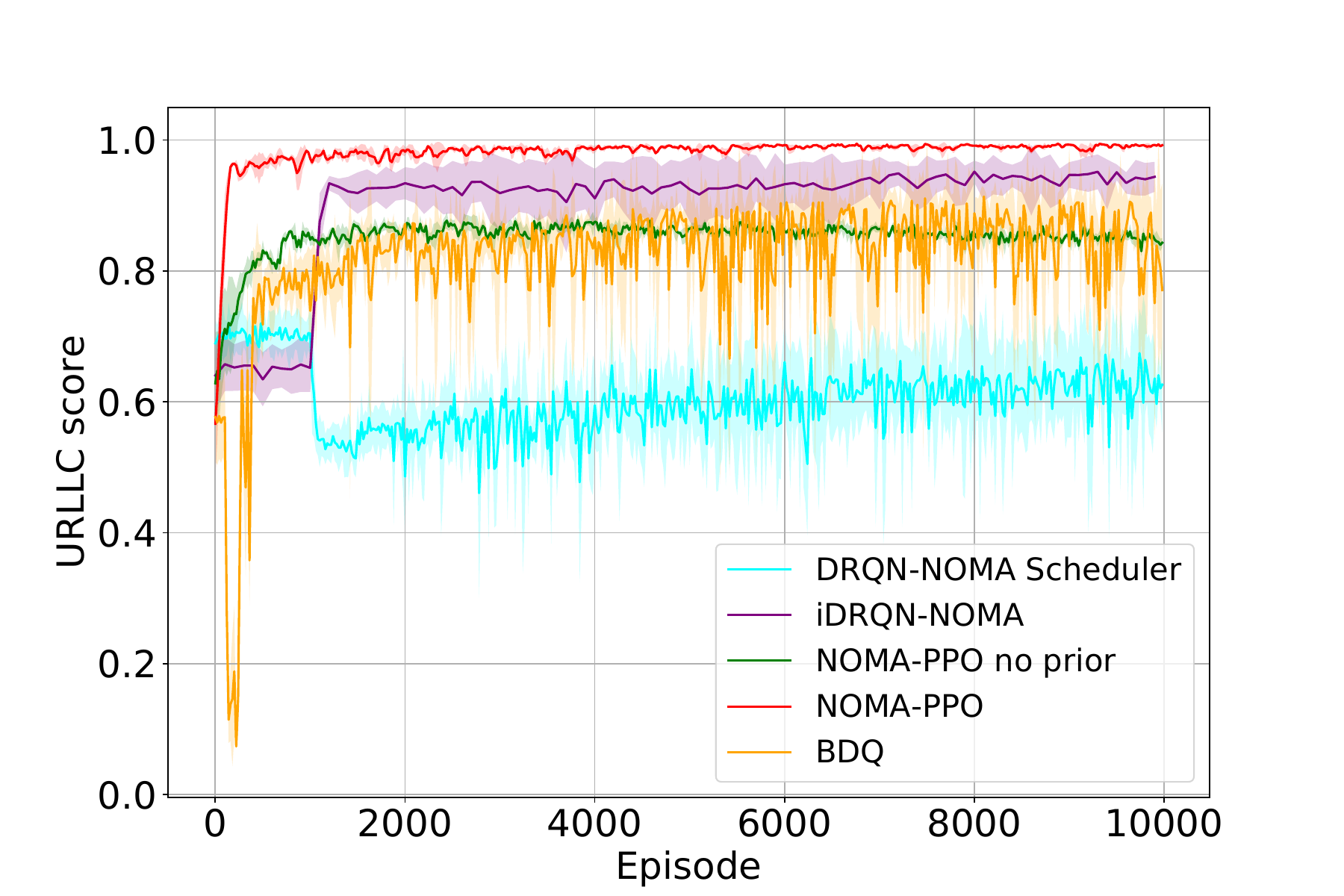}
        \caption{Evolution of the URLLC score during training for 18 users.}
        \label{fig:training}
    \end{subfigure}
    \hfill % it is used to fill in the space between the subfigures
    \begin{subfigure}[b]{0.48\textwidth} 
     \centering 
        \includegraphics[width=\textwidth]{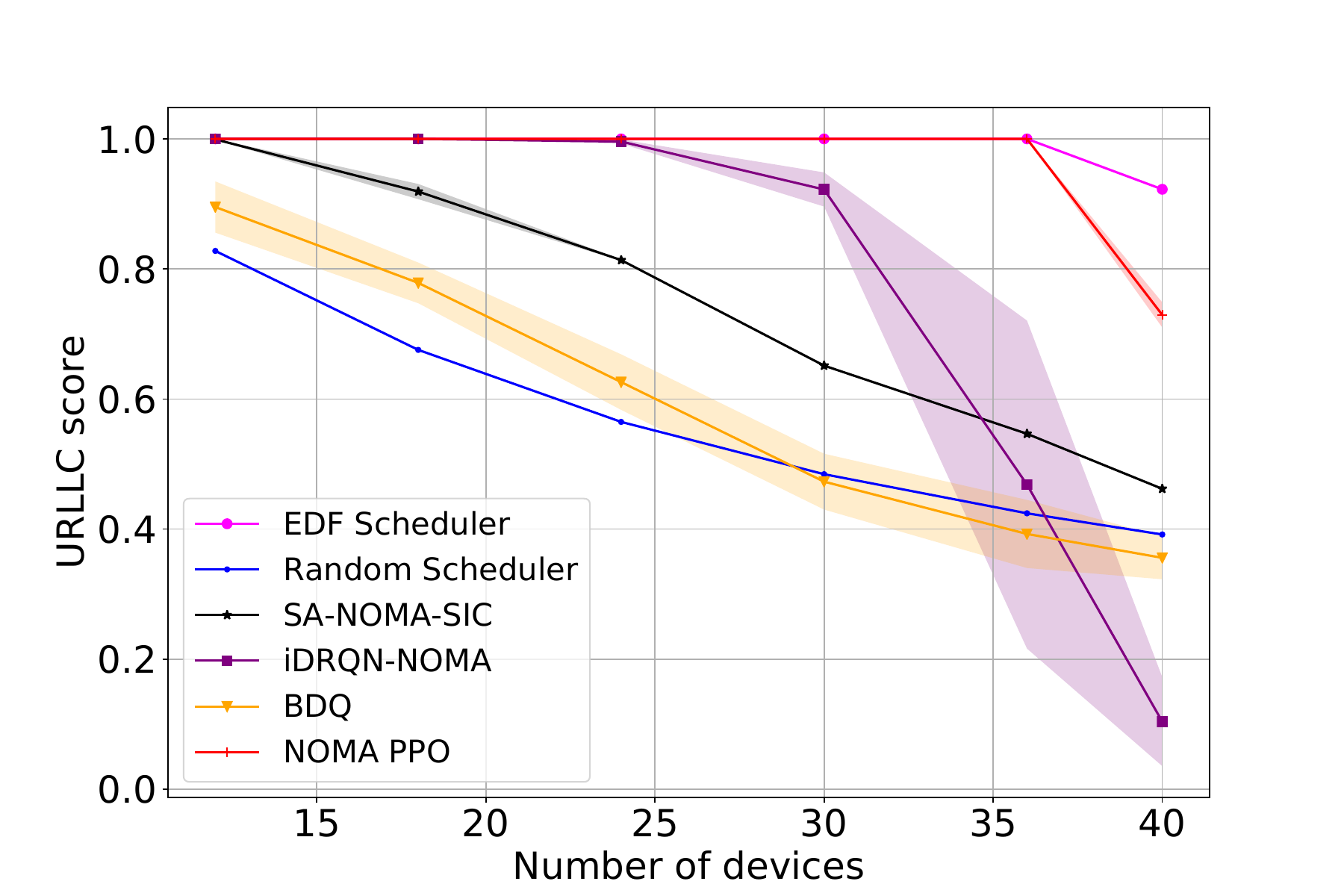}
        \caption{URLLC score in the 3GPP deterministic periodic scenario.}
        \label{fig:urllc-3gpp-periodic}
        
    \end{subfigure}

    \begin{subfigure}[b]{0.48\textwidth}   
        \centering
        \includegraphics[width=\textwidth]{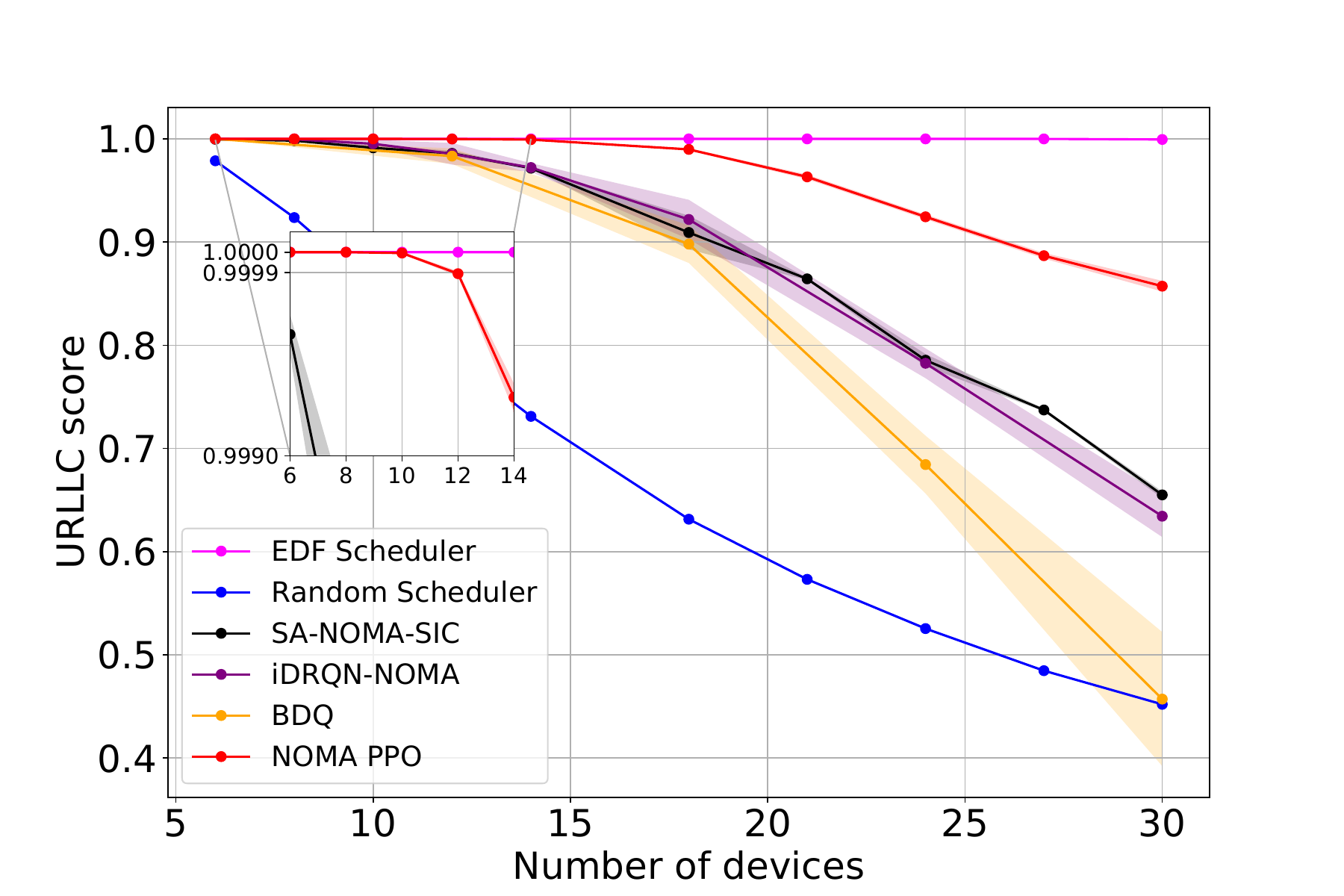}
        \caption{URLLC score in the 3GPP probabilistic aperiodic scenario.}
        \label{fig:urllc-3gpp-aperiodic}

    \end{subfigure}
    \hfill
    \begin{subfigure}[b]{0.48\textwidth}   
        \centering 
        \includegraphics[width=\textwidth]
        {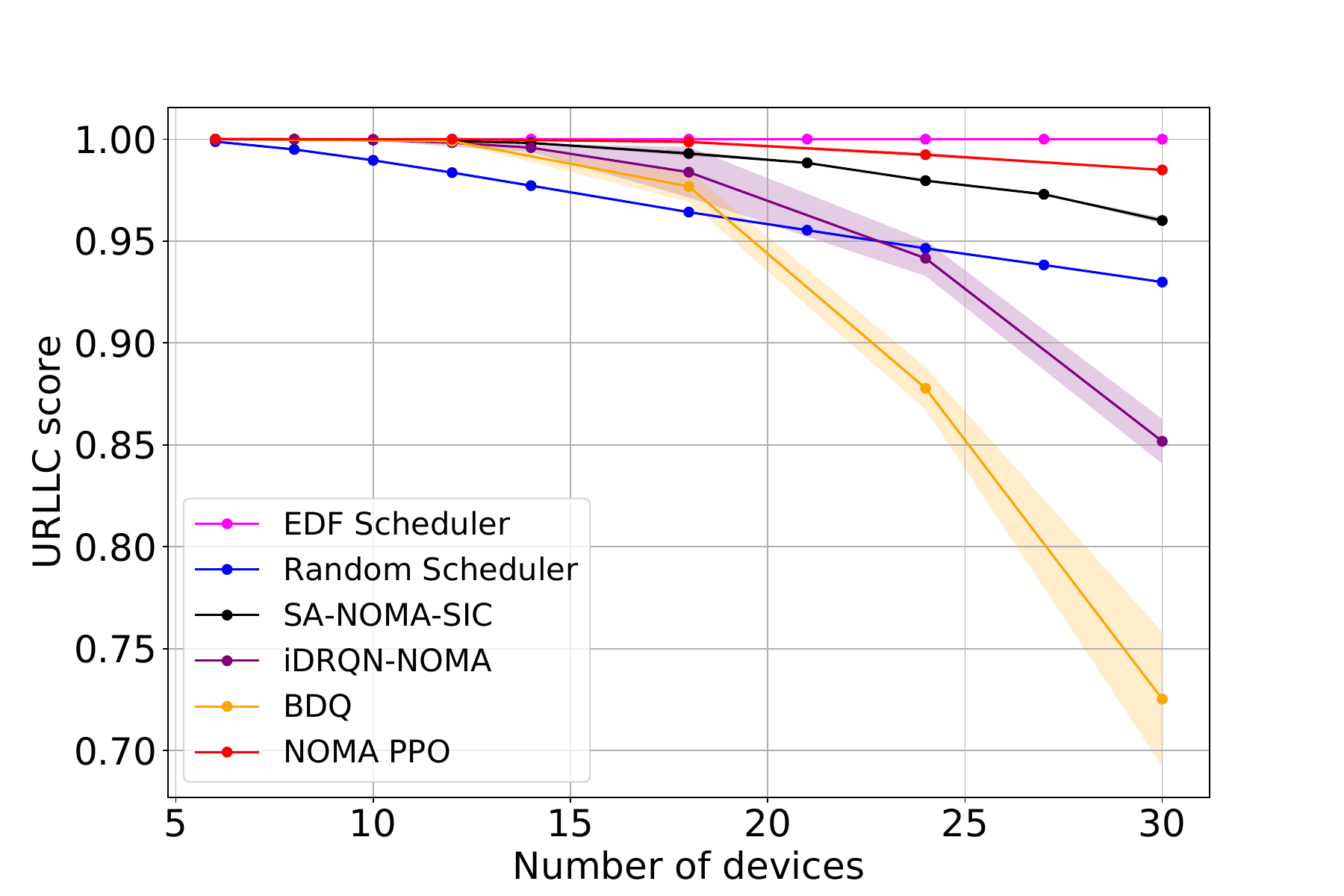}
        \caption{Jain's fairness index in the 3GPP probabilistic aperiodic scenario.}
        \label{fig:jains-3gpp-aperiodic}
    \end{subfigure}
    \caption{Performance metrics in the 3GPP scenario.}
    \label{fig:3gpp_scenario}
\end{figure*}

In Fig.~\ref{fig:3gpp_scenario}, we study the performance of our algorithm in the 3GPP scenarios with two different traffic models. On the one hand, we study in Fig.~\ref{fig:urllc-3gpp-periodic} the evolution of the URLLC score as a function of the number of devices on the deterministic periodic traffic and on the other hand, the evolution of the URLLC score and the Jain's Index on the probabilistic aperiodic traffic in Fig.~\ref{fig:urllc-3gpp-aperiodic} and Fig.~\ref{fig:jains-3gpp-aperiodic} respectively. The Jain's index is computed with the URLLC scores.

We observe that our approach, NOMA-PPO, outperforms all benchmarks in terms of URLLC score and fairness in all scenarios, except the EDF scheduler that has full observability over the devices' buffers. We notice that NOMA-PPO manages to handle the large action space better than the BDQ algorithm which fails to converge as we increase the number of devices. One explanation can be the fact that BDQ is not adapted to handle partial observability, a challenge that escalates as we increase the number of users. Regarding the grant free approaches, we first notice that iDRQN fails to converge for a number of devices greater than 30 in the deterministic periodic traffic in Fig.~\ref{fig:urllc-3gpp-periodic}. Yet, in the probabilistic aperiodic traffic, the MARL approach is slightly better than SA-NOMA-SIC for a number of users up to 18 and then becomes worse as the number of devices increases. This significant drop in performance can be explained by the limitations of independent learning, which creates instabilities during training due to the non-stationarity generated by the concurrent learning of the RL agents. %Another reason can be the fact that users and homogeneous which leaves the learning algorithm with less elements to take advantage of.

\subsection{Performance in Different Channel Conditions}

In this subsection, we analyze the behavior of NOMA-PPO under diverse channel conditions. 

\begin{figure*}[t] % the [t] aligns the figure at the top of the page
    \centering
    \begin{subfigure}[b]{0.49\textwidth} 
        \centering
        \includegraphics[width=\textwidth]{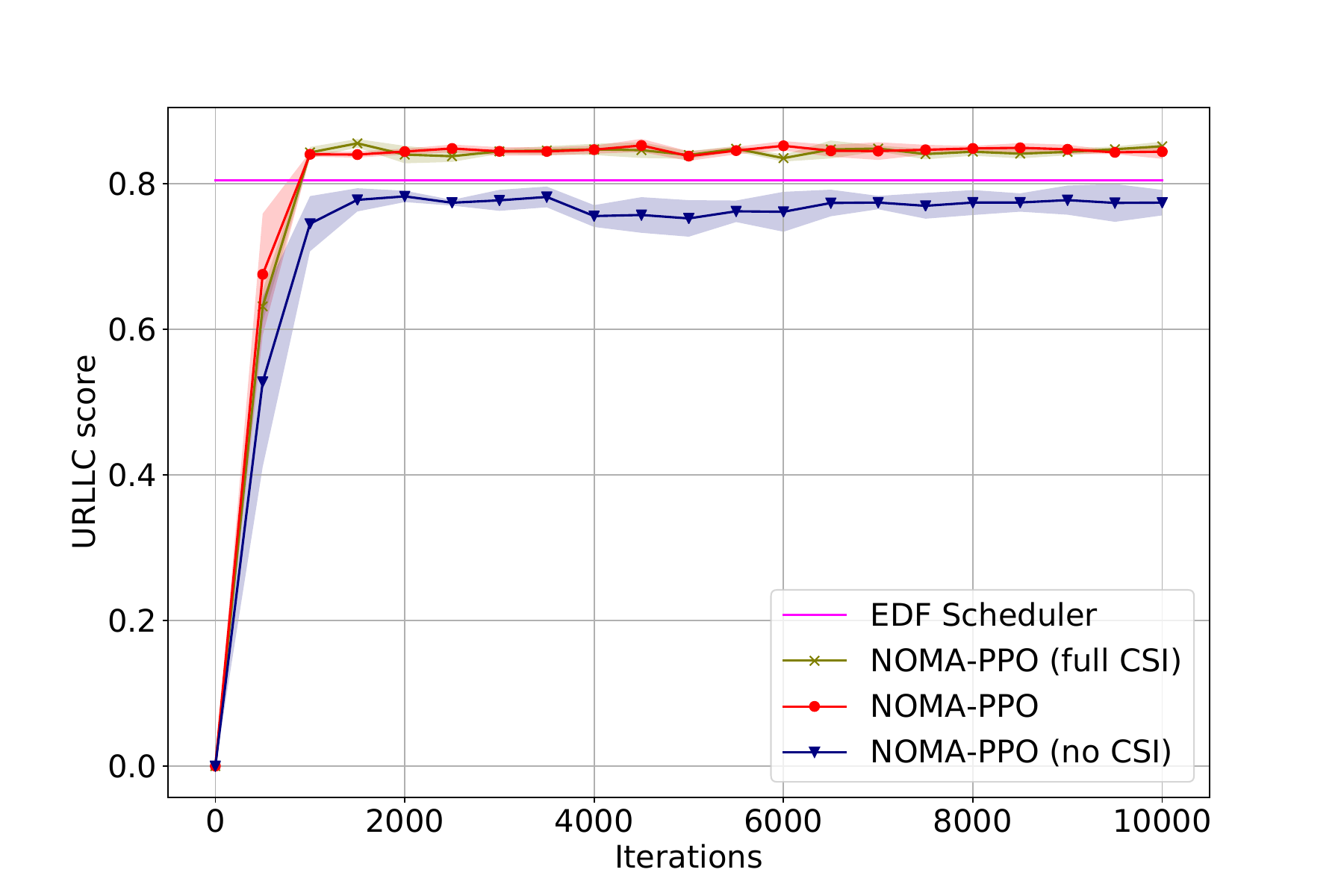}
        \caption{$T_c=1.4$ms, 10 users.}
        \label{fig:channel_ct=1.4}
    \end{subfigure}
    \hfill % it is used to fill in the space between the subfigures
    \begin{subfigure}[b]{0.49\textwidth}   
        \centering 
        \includegraphics[width=\textwidth]{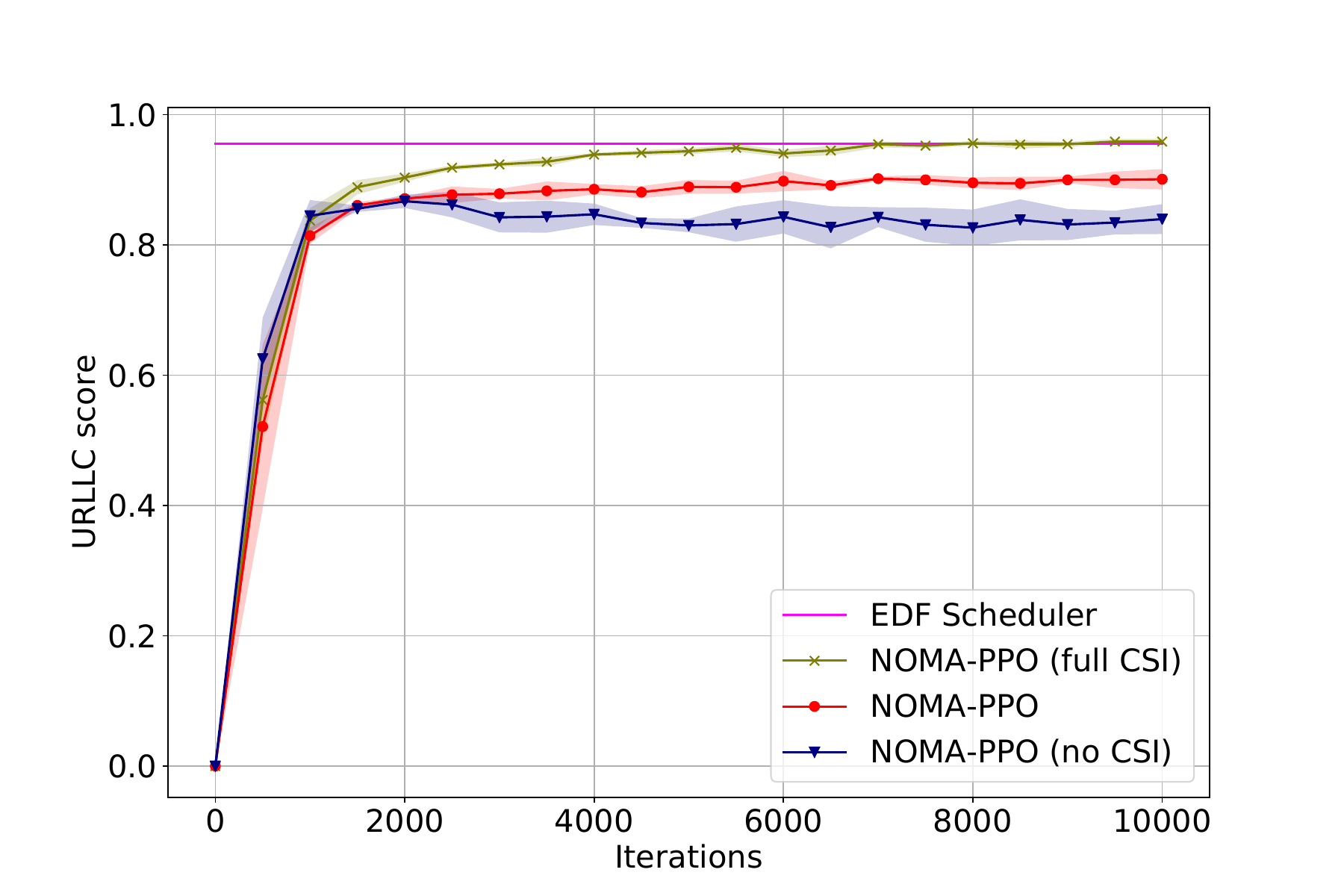}
        \caption{$T_c=0.34$ms, 10 users.}
        \label{fig:channel_xp_ct_fast}
    \end{subfigure}
    \caption{Evolution of the URLLC score during training under different channel conditions.}
    \label{fig:channel_experiments}
\end{figure*}

Fig.~\ref{fig:channel_experiments} shows the evolution of the URLLC score during the training as a function of the number of iterations in the probabilistic aperiodic traffic of the 3GPP scenario where parameters are listed in Table.~\ref{tab:simparam}, for 10 devices. Besides, we test two different values for the deadline-coherence time ratio where packets have a deadline of $10$~ms, a coherence time of $1.4$~ms (Fig.~\ref{fig:channel_ct=1.4}) and $0.34$~ms (Fig.~\ref{fig:channel_xp_ct_fast}). 

We compare the NOMA-PPO agent with:
\begin{itemize}
    \item \textbf{NOMA-PPO (full CSI)}: the version of NOMA-PPO where the channel information of all users is observable.
    \item \textbf{NOMA-PPO (no CSI)}: the version of the algorithm where we remove the channel information from the agent state.
\end{itemize}

First, we can see on all figures that the complete observability of the users' channels enriches the agent state and results in superior performance compared to the versions lacking this feature. Second, as depicted in Fig.~\ref{fig:channel_ct=1.4}, when the coherence time is long enough, NOMA-PPO manages to accurately estimate the true channel based on the agent state. This enables it to reach similar performance to NOMA-PPO with full CSI. Third, we observe in Fig.~\ref{fig:channel_xp_ct_fast}, when the coherence time is too short, that NOMA-PPO fails to reach the performance of NOMA-PPO with full observability but still outperforms the one with no CSI. The reason is that the coherence time must be large enough compared to the deadline so that the algorithm has time to both sense the channel and schedule the packet when the channel conditions are favorable. Furthermore, having the channel information in the agent state allows the agent to outperform the EDF scheduler that only schedules packets according to their deadline. 

Finally, we noted through simulations not depicted here, that when the number of users is too small, NOMA-PPO (no CSI) attains equivalent performance to both NOMA-PPO (full CSI) and NOMA-PPO. This can be explained by a multi-user diversity gain that increases with the number of devices.

\subsection{Complexity Analysis}

We evaluate the complexity of our Deep Learning architecture in terms of Floating Point Operations (FLOPs) that occur during a single forward pass of the neural network. Given that a connection between 2 neurons involves 2 operations (a multiplication and an addition) the FLOPs of a linear layer operation is $2\times \text{input size}\times \text{output size}$. The GRU layer is made of four operations: the reset gate, the update gate, the candidate hidden state and the new gate \cite{chung2014empirical}, each being made of matrix multiplications, additions, and activation functions. Let $H_{in}$ the size of the input and $H$ the size of the hidden layer. We can express the FLOPs of each unit:
\begin{itemize}
    \item The reset and update gate have the same structure: 2 matrix multiplications, 3 additions and 1 sigmoid activation. The resulting number of FLOPs for both gates is thus: $4H (H_{in} + H) + 6 H$.
    \item The new gate contribution includes 2 matrix multiplications, a Hadamard product, a $tanh$ activation and the addition of the bias terms: $2H(H_{in} + H) + 4H$.
    \item Finally, the new hidden state involves two Hadamard products, and two additions: $4H$.
\end{itemize}

In addition, the complexity of the GRU depends on the size of the history. In our problem, we set the history size to the number of users $K$. In total, the number of FLOPs of a GRU layer is $6HK(H_{in}+H)+10HK$.

\begin{table}[H]
\caption{FLOPs for the Deep Neural Networks.}
\centering
\begin{tabular}{|c|c|}
\hline
Algorithm & FLOPs \\
\hline
NOMA-PPO & $3,072K+263,424$ \\
BDQ & $4,096K+394,496$ \\
iDRQN-NOMA (1 agent) & $406,528K$\\
\hline
\end{tabular}
\label{tab:flops}
\end{table}

The number of FLOPs of the learning algorithms are given in Table~\ref{tab:flops} (using the numerical values of Table~\ref{tab:parameters_rl}). First, we can see that all algorithms have a linear complexity in the number of users and differ by their slope. Second, the BDQ algorithm has a complexity greater than NOMA-PPO due to the use of an advantage and a value network during inference. Third, the complexity of one iDRQN agent is larger than the complexity of the centralized approach we propose due to the use of a GRU layer to process the action-observation history.

\section{Conclusion}
In this paper, we propose a novel approach for satisfying Ultra Reliable Low Latency Communications (URLLC) requirements and strict deadlines in IoT networks, employing Non-Orthogonal Multiple Access (NOMA) for uplink communications. 

Our proposed approach, NOMA-PPO, addresses the challenges posed by the NOMA uplink URLLC scheduling problem, namely the combinatorial action space and the partial observability.
NOMA-PPO tackles these challenges by bringing three technical contributions. First, it formulates the NOMA-URLLC problem as a Partially Observable Markov Decision Process (POMDP), and introduces the concept of {\it agent state}, as sufficient statistic for the past actions and observations. This reformulation allows us to extend the state-of-the-art Proximal Policy Optimization (PPO) algorithm to handle a combinatorial action space thanks to a branching policy network. Finally, NOMA-PPO is able to incorporate prior knowledge over the system into the learning algorithm by employing a Bayesian policy.
We demonstrate that our approach outperforms traditional Multiple Access and Deep Reinforcement Learning benchmarks in 3GPP scenarios for different traffic models (probabilistic aperiodic and deterministic periodic) in terms of URLLC score, fairness, and convergence speed. Finally, we show that our algorithm is robust under diverse channel configurations and is capable to leverage channel information.

\appendices

\section{Proof of Proposition~\ref{prop:sufficient}} \label{app:sufficient}

    Let $\boldsymbol{s(t)}= \langle \boldsymbol{B}(t), \boldsymbol{\eta}(t), \boldsymbol{o}(t) \rangle$. We need to prove that $\boldsymbol{A(t)}$ is a sufficient statistic for the history $\hbar(t)$ in order to predict $\boldsymbol{s(t)}$ i.e.: 
    \begin{equation}
        P(\boldsymbol{s(t)} | \hbar(t)) = P(\boldsymbol{s(t)} | \boldsymbol{A(t)})
    \end{equation}  

According to the Bayes rule, we have: 

    \begin{align*}
        P(\boldsymbol{s}(t) | \hbar(t)) 
        &= P(\boldsymbol{B}(t), \boldsymbol{\eta}(t), \boldsymbol{o}(t) | \hbar(t)) \\
        &= P(\boldsymbol{B}(t)|\boldsymbol{\eta}(t), \boldsymbol{o}(t), \hbar_t) P(\boldsymbol{\eta}(t) | \boldsymbol{o}(t), \hbar(t)) P(\boldsymbol{o}(t)|\hbar(t))
\end{align*}

Besides, as $\hbar(t)=\{a(0),o(1),a(1),...,a(t-1),o(t)\}$ and $\boldsymbol{B}(t)$ is independent of $\boldsymbol{\eta}(t)$, we have:
    \begin{equation*}
        P(\boldsymbol{s}(t) | \hbar(t)) =  P(\boldsymbol{B}(t)|\hbar(t))P(\boldsymbol{\eta}(t) |\hbar(t))
    \end{equation*}

Each channel is independent so let's compute $P(\eta_{k}(t)|\hbar(t))$. For a device $k$, $\eta_k(t)$ is conditionally independent of $\boldsymbol{a}_k(t), \boldsymbol{b}^o_k(t)$ and $r(t-1)$ given $\eta_k^o(t)$. Thus, we can write:
    \begin{align*}
        P(\eta_k(t)|\hbar(t)) 
        &= P(\eta_k(t) | \eta_k^o(t-1), \dots, \eta_k^o(0)) \\
        & \stackrel{(a)}{=} P(\eta_k(t) | u_{k}(t-1) \eta_k(t-1), \dots, u_{k}(0)\eta_k(0)) \\
        &\stackrel{(b)}{=} P(\eta_k(t)|\eta_k(\tau^{a}_{k}(t)), \tau^{a}_{k}(t))\\
        &\stackrel{(c)}{=} P(\eta_k(t)| \boldsymbol{A}(t))
    \end{align*}
\noindent where (a) comes from the definition of $\eta^o_{k}$ (see Section~\ref{section:interference}-2); (b) comes from the fact that the channel realizations are Markovian and that $\eta_k(\tau^{a}_{k}(t))$ is the last observed channel realization for the device $k$; (c) results from the conditional independence of $\eta_k(t)$ from $\boldsymbol{B}^A(t), \boldsymbol{\tau}^a(t), \boldsymbol{\tau}^s(t), r(t-1)$ given $\eta_k^A(t)$ and $\tau^a_k(t)$. 

Finally, as each user's buffer is independent, we are going to prove by induction that $P(\boldsymbol{b}_{k}(t) | \hbar(t)) = P(\boldsymbol{b}_{k}(t) | \boldsymbol{A}(t)), \forall k, \forall t \geq 0$. First, $P(\boldsymbol{s}(0) | \hbar(0)) = P(s(0) | o(0)) = P(s(0) | \boldsymbol{A}(0))$. Let $t\geq 0$ and $k\in [1, K]$. Let's assume that $P(\boldsymbol{b}_{k}(t) | \hbar(t)) = P(\boldsymbol{b}_{k}(t) | \boldsymbol{A}(t))$.
By definition of the agent state (see Definition~\ref{def:agentstate}) and as $\boldsymbol{b}_{k}(t+1)$ only depends on $\boldsymbol{b}_{k}^A(t+1)$, we have:
\begin{equation}
    P(\boldsymbol{b}_{k}(t+1) | \boldsymbol{A}(t+1)) = P(\boldsymbol{b}_{k}(t+1) | \boldsymbol{b}_k^A(t+1))  
\end{equation}

\begin{equation}
    \boldsymbol{b}^A_k(t+1) = \left\{
    \begin{array}{ll}
        \boldsymbol{b}^{o}_k(t+1) & \mbox{if } \phi_{k}(t) = 1 \\
        \boldsymbol{b}^{A}_k(t) & \text{if } \phi_{k}(t) = 0
    \end{array}
\right.
\end{equation}

Therefore, 
\begin{align}\label{proof_buff}
    P(\boldsymbol{b}_{k}&(t+1) | \boldsymbol{b}_k^A(t+1))  = \begin{cases}
        P(\boldsymbol{b}_{k}(t+1) | \boldsymbol{b}^{o}_k(t+1)), & \text{if } \phi_{k}(t) = 1, \\
        P(\boldsymbol{b}_{k}(t+1) | \boldsymbol{b}^{A}_k(t)), & \text{if } \phi_{k}(t) = 0,
    \end{cases}
\end{align}

Besides, as $\boldsymbol{b}_{k}(t+1)$ is conditionally independent of $r(t)$, $\boldsymbol{H}^o(t+1), \boldsymbol{u}(t)$ given $\boldsymbol{\phi}(t), \boldsymbol{b}_{k}^o(t+1)$ we can write $P(\boldsymbol{b}_{k}(t+1) | \boldsymbol{b}_{k}^o(t+1)) = P(\boldsymbol{b}_{k}(t+1) | \boldsymbol{o}(t+1))$.
Therefore:
\begin{align*}
    P(\boldsymbol{b}_{k}(t+1) | \boldsymbol{b}_k^A(t+1)) &\stackrel{(c)}{=} P(\boldsymbol{b}_{k}(t+1) | \boldsymbol{b}^A_k(t), \boldsymbol{o}(t+1)) \\ 
    &  \stackrel{(d)}{=} P(\boldsymbol{b}_{k}(t+1) | \boldsymbol{A}(t), \boldsymbol{a}(t), \boldsymbol{o}(t+1))\\
    & \stackrel{(e)}{=} P(\boldsymbol{b}_{k}(t+1) | \hbar(t+1))
\end{align*}

\noindent where (c) comes from merging the equations in (\ref{proof_buff}) and noticing that $\phi_k(t) \in \boldsymbol{o}(t+1)$; (d) comes from the fact that $\boldsymbol{b}_{k}(t+1)$ only depends on $\boldsymbol{A}(t), \boldsymbol{a}(t), \boldsymbol{o}(t+1)$ through $\boldsymbol{b}^A_k(t)$ and $\boldsymbol{o}(t+1)$. Finally, (e) comes from the induction hypothesis.

\bibliographystyle{IEEEtran}

\bibliography{references/3gppspec36, references/3gppspec37, references/3gppspec38, references/references}

\end{document}